\documentclass[12pt,preprint]{aastex}

\usepackage{amsmath}

\newcommand\cmt{{\rm cm^{-3}}}

\newcommand\Msun{{\rm\,M_\odot}}

\newcommand\kms{{\rm km\, s^{-1}}}
\newcommand\pc{{\rm\,pc}}

\newcommand\simgt{\lower.5ex\hbox{$\; \buildrel > \over \sim \;$}}
\newcommand\simlt{\lower.5ex\hbox{$\; \buildrel < \over \sim \;$}}

\shorttitle{Thermal and Magnetorotational Instabilities in the ISM}
\shortauthors{Piontek \& Ostriker}

\begin{document}


\title{Saturated-State Turbulence and Structure from Thermal and
  Magnetorotational Instability in the ISM: Three-Dimensional
  Numerical Simulations}


\author{Robert A. Piontek and Eve C. Ostriker}
\affil{Department of Astronomy\\University of Maryland\\
    College Park, MD  20742-2421}
\email{rpiontek@astro.umd.edu, ostriker@astro.umd.edu}



\begin{abstract}  

This paper reports on three-dimensional numerical
simulations of dynamics and thermodynamics in the diffuse interstellar
medium (ISM).  Our models are local, account for sheared galactic
rotation, magnetic fields, and realistic cooling,
and resolve scales $\approx 1-200$ pc.  This combination permits the
study of quasi-steady-state turbulence in a cloudy medium representing
the warm/cold atomic ISM. Turbulence is driven by the
magnetorotational instability (MRI); our models are the first to
study the saturated state of MRI under strongly inhomogeneous
conditions, with cloud/intercloud density and temperature contrasts
$\sim 100$.  For volume-averaged densities $\bar n=0.25 - 4 \ \cmt$, the
mean saturated-state velocity dispersion ranges from $8-1 \ \kms$, with
a scaling $\delta v\propto \bar n^{-0.77}$. The MRI is therefore
likely quite important in driving turbulence in low-density regions of
the ISM, both away from the midplane in the inner Galaxy (as observed
at high latitudes), and throughout the far outer Galaxy (where the
mean density drops and the disk flares).  The MRI may even be key to
suppressing star formation at large radii in spiral galaxies, where
the pressure can be high enough that without MRI-driven turbulence, a
gravitationally-unstable cold layer would form.  As expected, we find
that turbulence affects the thermal structure of the ISM. In all our
simulations, the fraction of thermally-unstable gas increases as the
MRI develops, and in the saturated state is largest in high-$\delta v$
models.  The mass fractions of warm-stable and unstable gas are
typically comparable, in agreement with observations. While inclusion
of resistive dissipation of magnetic fields could enhance the amount
of thermally-unstable gas compared to current models, our present
results indicate that even high levels of turbulence cannot wipe out
the signature of thermal instability, and that a shift to a ``phase
continuum'' description is probably unwarranted.  Instead, we find
that temperature and density PDFs are broadened (and include extreme
departures from equilibrium), but retain the bimodal character of the
classical two-phase description.  Our presentation also includes
results on the distribution of clump masses (the mass spectrum peaks
at $\sim 100 \ \Msun$), comparisons of saturated-state MRI scalings with
single-phase simulation results (we find $\langle B^2 \rangle$ is
independent of $\bar n$), and examples of synthetic HI line profile
maps (showing that physical clumps are not easily distinguished in
velocity components, and vice versa).

\end{abstract}




\section{Introduction}

Far from the energizing regions of star formation in the Milky Way and
other galaxies, the interstellar medium (ISM) is still roiling with
activity, and rife with structure.  Both the microphysical properties
and turbulent activity have been increasingly well characterized by
Galactic and extragalactic radio observations.  In particular, recent
high-resolution Galactic emission surveys in the 21 cm hydrogen line
(e.g. \citet{mcc01, tay03}), combined with Galactic absorption surveys
(e.g. \citet{hei03, moh04}), and mapping of face-on external galaxies
(e.g. \citet{dic90a, van99}), have begun to provide a wealth of thermal
and kinematic information about the atomic ISM component, which
comprises the majority of the total ISM mass in most spiral galaxies.
Analysis of this data promises to yield a detailed empirical
description of the atomic gas, which is known to consist of both warm
and cold components, and to be strongly turbulent (e.g.
\citet{dic90b}).

As observations of the ISM advance, there is a need on the theoretical
side for increasingly sophisticated ISM modeling.  With modern
computational tools, it is possible to pursue time-dependent
hydrodynamic models which incorporate many physical processes.  This
numerical modeling can extend established ``classical'' results for
simplified systems into more realistic regimes, and test conceptual
proposals for the behavior of complex systems in a rigorous fashion.
The goal of detailed ISM modeling, of course, is not sophistication
for its own sake, but to gain understanding about how different
``elemental'' processes interact, to ascertain which among many
contributing processes are most important, and to aid in interpreting
and developing reliable physical diagnostics from observations.

Broadly, the presence of structure in the atomic ISM can be easily
understood as a consequence of the bistable thermal equilibrium curve
over a range of pressures, including those typical of the ISM.  Since
the temperatures of the two stable thermal equilibria differ by a
factor of $\sim 100$ (at fixed pressure), the ``classical''
expectation based on the principle of pressure equilibrium is a system
of cold, dense clouds embedded in a much more diffuse warm intercloud
medium \citep{fie69}.  Thermal instability (TI) tends to move gas
parcels at intermediate temperatures into one of the stable phases
\citep{fie65}.  Clouds are initially expected to condense at preferred
scales where conduction limits local thermal gradients.  While these
basic processes are certainly involved in establishing the ISM's
structure, the end result is a complex product of evolution and
interactions with other physical processes, leaving many open
questions.  For example, how do the agglomerations and disruptions of
cold clouds depend on the turbulence properties, and how does this
affect the mass function of condensations that results?

Many processes have been proposed that can produce turbulence in the
ISM (see e.g. \citet{mac04,elm04} for recent reviews).  Traditionally,
turbulence is thought to be driven primarily by supernovae
\citep{mck77} (and, to a lesser extent, expanding HII regions),
because the total kinetic energy they are able to supply could be
sufficient to offset the turbulent dissipation in the Milky Way's
diffuse ISM (Spitzer 1978, Ch. 11). Supernovae are certainly the
primary source of turbulence near regions of high-mass star formation.
However, it is not clear how effectively this energy can in fact be
shared with the bulk of the ISM, so other sources may be (or combine
to be) of comparable importance. Indeed, observations indicate that
the levels of turbulence are not strongly correlated with spiral arms
(where star formation is enhanced), and are just as large in outer
galaxies (where overall star formation rates are low) as in inner
regions \citep{van99, pet01}.  Moreover, recent 3D simulations
\citep{kor99,dea05} in which turbulence is driven solely by supernovae find
that velocity dispersions are significantly lower in cold gas than in
warm gas, inconsistent with observations \citep{hei03}.

An obvious non-stellar energy source for the ISM is galactic rotation.
Wherever the angular velocity decreases outward and magnetic fields
are present, the powerful magnetorotational instability (MRI) is
expected to tap this rotation and drive large-amplitude ISM turbulence
\citep{sel99,kim03,dzi04}.  Detailed development of MRI has primarily
been studied in adiabatic or isothermal gas, where turbulent
velocities and Alfv\'en speeds grow into rough equipartition at
slightly subsonic levels (e.g. \cite{haw95a, haw96} hereafter HGB1,
HGB2)).  Adiabatic and isothermal models, however, are essentially
single phase, with only small variations in density and temperature.
How do turbulent saturation levels differ in a medium where there are
huge variations in conditions, such that subsonic speeds with respect
to the diffuse gas are highly supersonic with respect to the dense
gas?

In the real ISM, dynamics must affect thermodynamics, and vice versa.
The turbulent power input is significant, and both (irreversible)
dissipative heating and (reversible) PdV heating and cooling can alter
distributions of temperatures compared to the narrow spikes at warm
and cold equilibria that would otherwise occur.  In turn,
thermodynamics potentially can affect loss rates of turbulence:
supersonic compressions are dissipative while subsonic compressions
are not, and dissipation of magnetic energy by reconnection depends on
local conditions of density and temperature. Cloudy structure also
changes effective flow ``collision'' times, as well as field line
geometry.  Indeed, recent observational evidence has shown that the
fraction of unstable gas in the ISM may be significant; \citet{hei03}
found that at high latitudes, about half the warm neutral medium (WNM)
lies at thermally unstable temperatures between 500-5000 K.  Numerical
models which include effects of star formation
\citep{ros95,kor99,dea00,wad00,gaz01,wad01a,wad01b,mac04,sly04} find both
turbulence and significant fractions of unstable gas, although it is
not clear how much the temperature distributions are affected by the
direct heat inputs in the star formation feedback algorithms of these
models.

Recent simulations have addressed nonlinear evolution, in 2D and 3D,
of TI in the ISM without ``stellar'' energy inputs
\citep{hen99,bur00,vaz00,san02,kri02,vaz03,aud04,kri04}, and there have
also been many numerical studies, in 2D and 3D, of the MRI in
single-phase gas.  In previous work, we performed 2D studies of TI and
MRI in combination (\cite{pio04}, hereafter Paper I).  Paper I showed
that MRI growth rates in a two-phase medium are comparable to those in
a single-phase medium with the same $\bar\rho$ and $\bar{\bf B}$,
provided that the cloud separation along field lines does not exceed
half of the fastest-growing MRI wavelength (typically $\sim 100 \pc$).
Although there have been suggestions that TI itself could be a
significant source of turbulence, ``pure TI'' models we performed show
that for pressures comparable to mean galactic values (i.e. away from
HII regions or recent supernovae), velocity dispersions are only a few
tenths of a $\kms$.  In our 2D simulations, the MRI leads to
large-amplitude velocities and magnetic fields, but as for
single-phase 2D models, late time behavior is dominated by the
``channel flow;'' quasi-steady turbulence is possible only for 3D
flows.  The present work constitutes the extension of Paper I to 3D,
in order to study the saturated state of MRI in the presence of a
two-phase medium.  As we shall describe, we have performed a variety
of simulations, with parameters covering a range of conditions
characteristic of the atomic ISM.

The plan of this paper is as follows: In \S 2 we briefly describe the
numerical method, and the initializations for the various models we
have performed. In \S 3 we present the results of our simulations in
terms of the models' physical structure, thermodynamic distributions,
and turbulent states (in velocities and magnetic fields), as well as
exhibiting sample synthetic observations based on our simulated data.
We summarize, discuss the astronomical implications of our results,
and compare to previous work in \S 4.

\section{Numerical Methods and Model Parameters}
\label{numerics}

The numerical methods utilized for the present study are essentially
the same as those of Paper I, but extended from 2D to 3D.  For a
complete description of the numerical method and tests, please see
that work.  Here, we briefly summarize the salient points.

We integrate the time-dependent equations of magnetohydrodynamics with
a version of the ZEUS code \citep{sto92a,sto92b}.  ZEUS uses a
time-explicit, operator-split, finite difference method for solving
the MHD equations on a staggered mesh, capturing shocks via an
artificial viscosity.  Velocities and magnetic field vectors are
face-centered, while energy and mass density are volume-centered.
ZEUS employs the CT and MOC algorithms \citep{eva88, haw95b} to
maintain $\nabla \cdot {\bf B} = 0$ and ensure accurate propagation of
Alfv\'{e}n waves.
     
We have implemented volumetric heating and cooling terms, and a
thermal conduction term.  The update due to net cooling is solved
implicitly using Newton-Raphson iteration.  For a given hydrodynamical
time step, the change in temperature in each zone is limited to be
less than 25\%.  This is a somewhat larger fraction than the 10\%
limit used in Paper I, which allows us to run with larger time steps
needed to make 3D calculations practical.  Tests have shown that
relaxing this constraint does not affect cloud structure; $\Delta$T
exceeds 10\% only in a very small fraction of zones.  The conduction
term is solved explicitly using a seven point stencil for the second
derivative of temperature. We also model the differential rotation of
the background flow and the variation of the stellar/dark matter
gravitational potential in the local limit with $x\equiv R-R_0 \ll
R_0$, where $R_0$ is the galactocentric radius of the center of our
computational domain.  The equations we solve are therefore:

\begin{equation}
\frac{\partial\rho}{\partial t}+ \boldsymbol{\nabla} \cdot (\rho
{\bf v}) = 0
\end{equation}

\begin{equation}
\frac{\partial{\bf v}}{\partial t}+
{\bf v}\cdot\boldsymbol{\nabla}{\bf v}=-\frac{\boldsymbol\nabla P}{\rho} + 
\frac{1}{4\pi\rho}(\boldsymbol{\nabla} \times {\bf B})
\times {\bf B}+ 2 q \Omega^{2}x\hat{x}-2\boldsymbol{\Omega}\times 
{\bf v}
\label{mom}
\end{equation}

\begin{equation}
\frac{\partial \mathcal{E} } { \partial t }
+{ \bf v}\cdot\boldsymbol{\nabla}\mathcal{E} = 
-(\mathcal{E} + P)\boldsymbol{\nabla}\cdot{\bf v}-\rho\mathcal{L}+
\boldsymbol{\nabla}\cdot(\mathcal{K}\boldsymbol{\nabla}T)
\label{energy}
\end{equation}

\begin{equation}
\frac{\partial {\bf B}}{\partial t}=\boldsymbol{\nabla \times}({\bf v}
\times {\bf B})
\end{equation}

All symbols have their usual meanings.  The net cooling per unit mass
is given by $\mathcal{L}=\rho\Lambda(\rho,T)-\Gamma$.  We adopt the
simple atomic ISM heating and cooling prescriptions of \citet{san02},
in which the cooling function, $\Lambda(\rho,T)$, is a piecewise
power-law fit to the detailed models of \citet{wol95}.  The heating
rate, $\Gamma$, is taken to be constant at 0.015 $\rm{erg\ s^{-1}
  g^{-1}}$.  In the tidal potential term of equation (\ref{mom}),
$q\equiv-d \ln \Omega / d \ln R$ is the local dimensionless shear
parameter. We adopt $q$ equal to unity, to model a flat rotation curve
in which the angular velocity $\Omega \propto R^{-1}$.

The present set of simulations is 3D, with the computational domain
representing a cubic sector of the ISM in the
radial-azimuthal-vertical ($x-y-z$) domain.  We employ
shearing-periodic boundary conditions in the $\hat{x}$-direction
\citep{haw92,haw95a}, and periodic boundary conditions in the
$\hat{y}$- and $\hat{z}$-directions, as originally implemented in ZEUS
by \citet{sto96}.  This framework allows us to incorporate realistic
galactic shear, while avoiding numerical artifacts associated with
simpler boundary conditions.  We have previously used an isothermal
version of the same code to study larger-scale galactic ISM problems
\citep{kim02,kim03}.

We have parallelized the code with MPI to run on distributed-memory
platforms.  We perform a standard domain decomposition in $\hat{x}$,
so that each processor works on a slab of the domain.  Decomposing in
$\hat{y}$ and $\hat{z}$ as well would reduce message-passing time when
running on more than eight processors, but for gigabit networks and
faster the compute time exceeds the message-passing time for 128 or
fewer processors in any case. Thus, for our moderate problem sizes,
the additional effort that would be required to parallelize the
shearing-periodic boundary conditions is not merited.

Our standard resolution is $128^3$ zones.  To confirm numerical
convergence, we also performed one simulation at $256^3$ and found the
results to be similar to the standard resolution run.  Our box is
200pc on each side, giving a resolution of about $0.8 \ {\rm pc}$ in
our $256^3$ run, and about $1.6 \ {\rm pc}$ in our $128^3$ resolution
runs.  We set the conduction coefficient to $\mathcal{K}=1.03 \times
10^{7} \rm{ergs \ cm^{-1} \ K^{-1} \ s^{-1}}$.  This level of
conduction was chosen to allow us to resolve all modes of TI that are
present (see Paper I).  In our fiducial model, the initial state of
the gas is constant density, constant pressure with conditions
comparable to mean values in the solar neighborhood; $n=1.0 \ 
\rm{cm}^{-3}$ and $P_0/k=2000 \ \rm{K \ cm^{-3}}$.  The
corresponding initial isothermal sound speed is $c_s=3.6 \ {\rm km \ 
  s^{-1}}$. The initial magnetic field strength satisfies
$\beta=P_{gas}/P_{mag}=100$, corresponding to $B=0.26 \ \mu$G in the
vertical ($\hat{z}$) direction.  This is our ``standard'' run.
Additional simulations are performed with varying mean densities of
$n=0.25,0.67,1.5$ and 4.0 $\rm{cm^{-3}}$, as well as one with
lower magnetic field strength, $\beta=P_{gas}/P_{mag}=1000$.  We also
performed an isothermal simulation with $c_s=2.8 \ \rm{km \ s^{-1}}$
and $n=1.0 \ \rm{cm}^{-3}$.  This value of $c_s$ was chosen so that the initial
thermal pressure matches the mean late-time pressure in our cooling
models.  Finally, we also performed a simulation with heating and
cooling turned on that was initialized from the saturated-state,
turbulent isothermal model.  For all our models we adopt the galactic
orbital period at the solar radius, $2.5 \times 10^8 \ \rm{yr}$, to
normalize the shear rate.

Since increasing or decreasing the mean density by a large factor
relative to $n=1 \ \rm{cm^{-3}}$ would initialize the gas in a
thermally $stable$ state, some of our simulations are initialized with
a medium already in a two-phase state, rather than with a uniform
density.  For these models, spherical clouds of cold dense gas are
inserted into a warm ambient medium at random locations.  The number
of clouds is adjusted so that the average density of the cloudy medium
is at the desired level.  A similar simulation was performed in Paper
I, which allowed us to study the growth rates of the MRI in an
initially quiescent cloudy medium.  Since the 2D simulations of Paper
I were axisymmetric there was no evolution of the model until MRI
modes began to grow.  This allowed us to compare directly the MRI
growth rates of an adiabatic run with a two-phase run, illustrating
the effect of cloud size and distribution on the growth rates.  In the
present 3D simulations, however, the evolution is rapid because the
symmetry in the azimuthal direction is broken.  Individual clouds are
sheared out relatively quickly, and also begin to merge with nearby
clouds. Nevertheless, because MRI-driven turbulence eventually
dominates both the initially-thermally-unstable and
initially-two-phase models, at late times the two are
indistinguishable.

On top of the initial conditions given above, we add pressure
perturbations with a white noise spectrum at the 0.1\% level to seed
the TI and MRI.  In the next section, we describe results from our
standard run in detail, and comment on differences with the other runs
as is appropriate.

\section{Results}

\subsection{Overall Evolution}
\begin{figure}
\plotone{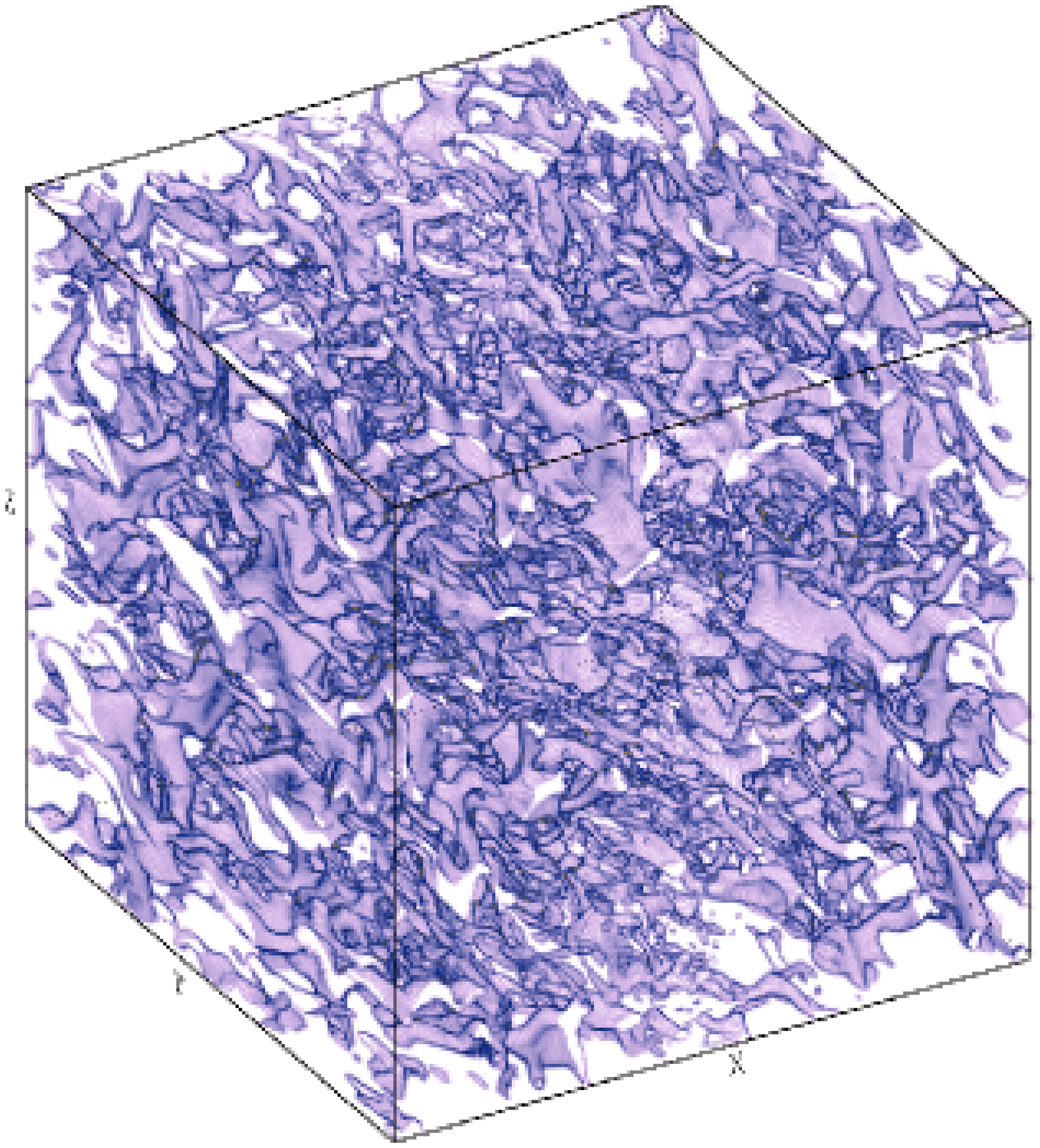}
\caption{Volume rendering of density for the standard run at t=1.0 orbits.  The ``y'' direction is azimuthal in this model, and the ``x'' direction is radial.  Most of the mass is in the cold phase, while most of the volume is occupied by the warm phase.
\label{vol1}}
\end{figure}

\begin{figure}
\plotone{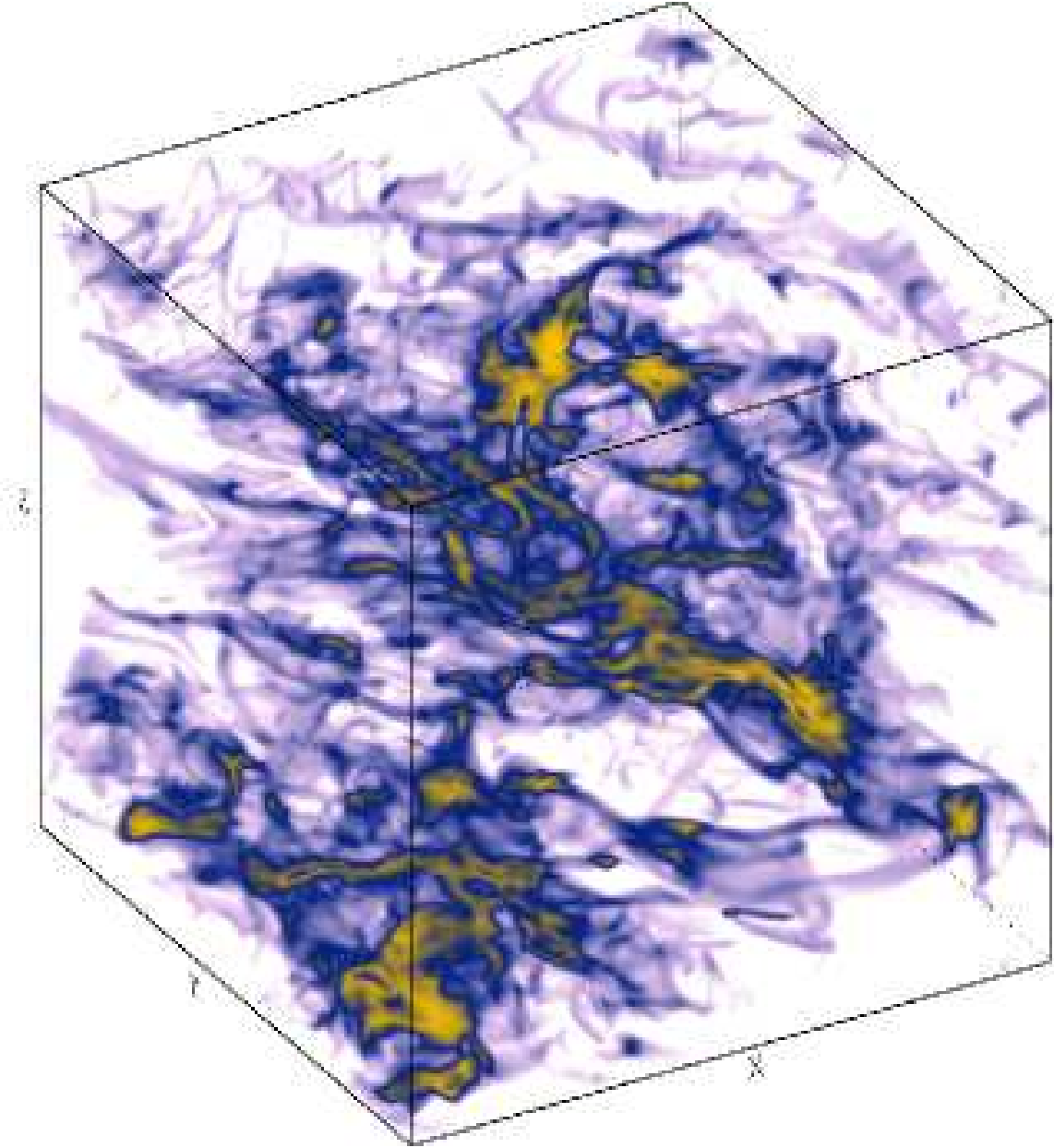}
\caption{Volume rendering of density for the standard run at t=9.0 orbits.  Turbulence due to the MRI has forced gas to higher densities, as well as increased the mass fraction of the unstable phase. Many filamentary structures are present.
\label{vol2}}
\end{figure}

Figures \ref{vol1} and \ref{vol2} are volume renderings of the 3D
density data cube, from our run with fiducial parameters, and
resolution $256^3$, at $t=1.0$ and $9.0$ orbits.  The early
development of both TI and MRI in the present set of 3D simulations is
quite similar to the development previously described for 2D
simulations in Paper I.  Initially the gas is thermally unstable.  The
cooling time scale is much shorter than the orbital time scale, and
the gas quickly separates into many small, cold clouds embedded in a
warm ambient medium.  This phase of the evolution lasts about 20 Myr,
which is comparable to the 2D simulations of Paper I.  The typical
size scale of the clouds is about 5 pc, consistent with expectations
for the fastest growing modes at the adopted level of conductivity.
The size scale of the clouds is still fairly close to its initial
distribution in Figure \ref{vol1} at $t=1.0$ orbits.

After the initial condensation phase of TI is complete, large scale
galactic shear begins to drive the evolution.  Already at $t=2.0$
orbits, the clouds have become elongated in the $\hat{y}$ direction.
During the first few orbits interactions take place between nearby
clouds, which typically lead to mergers, increasing the typical size
scale significantly.  At about $t=4.0$ orbits ($=10^9 \rm{yrs}$)
the modes of the MRI have grown
significantly and now begin to dominate the evolution of the model.
The simulation becomes fully turbulent, drastically altering the
dynamics compared to the axisymmetric model of Paper I.  Shear from
the MRI with velocities in all directions, combined
with galactic shear with velocities in the azimuthal direction, leads
to repeated disruptive interactions and collisions between clouds.
Clouds merge into an interconnected network, with individual entities
existing for only short periods of time.  It is difficult to convey
the dynamical nature of the simulations to the reader using only
snapshots in time; the animation associated with Figures
\ref{vol1}-\ref{vol2} shows this much more clearly.  

While the structure remains highly dynamic, a quasi-equilibrium
saturated state is established by $t \sim 5$ orbits, and the
statistical properties of the gas remain relatively constant
throughout the latter half of the simulation (up to t=10 orbits).  The
approach to a quasi-steady turbulent state in these models is
generally similar to the results for isothermal or adiabatic single
phase models (e.g. HGB1, HGB2).  In the remainder of \S 3, we discuss
details of evolution and quasi-steady properties, similarities and
differences from single-phase models, and dependencies on model
parameters.

\subsection{Density Structure}
\begin{figure}
\plotone{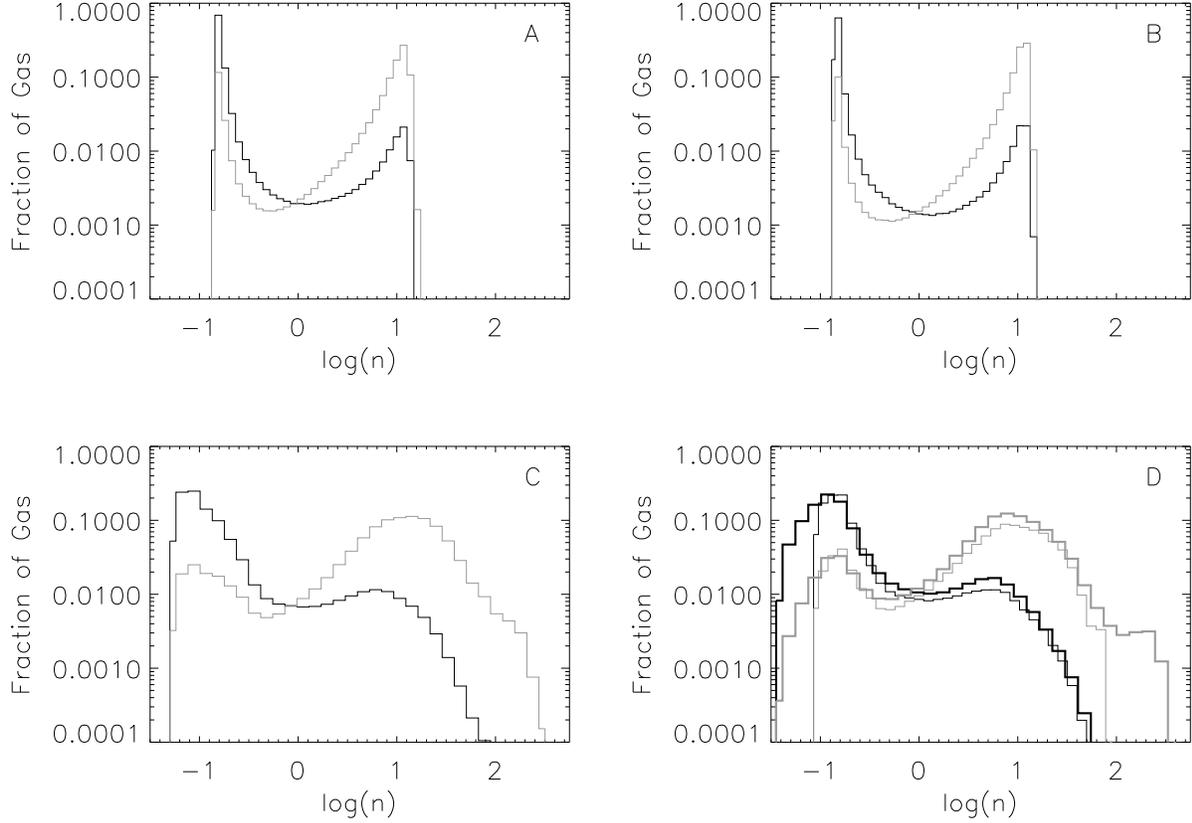}
\caption{Volume (black line) and mass (grey line) PDFs of density for the standard run  at $t=1, 2.5$, and $5.0$ orbits (panels A-C, respectively).  In Panel D we compare the PDFs for our two resolutions of $128^3$ (thin lines) and $256^3$ (thick lines) at $t=9$ orbits.  The PDF changes little between Panels A and B.  In Panels C and D the development of the MRI increases the fraction of gas in the unstable phase, and forces gas to higher densities.  In Panel D, the higher resolution run contains trace amounts of gas at more extreme densities. 
\label{denpdf}}
\end{figure}

The density probability distribution functions (PDFs) from our
standard run (at $128^3$) are shown in Figure \ref{denpdf} at $t=1,
2.5, 5.0,$ and $9$ orbits.  We show both mass-weighted and
volume-weighted density PDFs in Panels A-D, and compare the PDFs of
the $128^3$ and $256^3$ runs in Panel D.  Similar to our results in
Paper I, we find that by mass, most of the gas is in the cold phase,
while the warm phase occupies most of the volume.  After the initial
development of TI has completed, at $t=1.0$ orbits, the mass fraction
of gas in the warm (F), unstable (G), and cold (H) phases is 14\%,
5\%, and 80\%, respectively.  By volume, 83\%, 9\%, and 8\% of the gas
resides in the warm, unstable, and cold phases.  From $t=1.0$ to
$t=2.5$ orbits (panels A and B of Figure \ref{denpdf}) the evolution
is driven mainly by galactic shear.  The size distribution of the
clouds shifts to larger masses through mergers, but the density PDFs
over this interval vary little.  The fraction of gas in each phase
changes by only a few percent during this time period.

In contrast, between $t=2.5$ and $t=5.0$, Panels B and C of Figure
\ref{denpdf}, the evolution changes from being driven primarily by
galactic shear, to being driven primarily by the MRI.  The model
becomes fully turbulent, and this has a significant effect on the
detailed shape of the density PDF.  The fractions of gas in the warm,
unstable, and cold phases at $t=5.0$ are now 10\%, 7\%, and 83\% by
mass, and 84\%, 8\%, and 7\% by volume. Near the end of the
simulation, at $t=9$, the gas fractions are 14\%, 18\%, and 67\%
percent by mass and 82\%, 10\%, and 6\% percent by volume.  From $t=5$
to $t=9$, (Panel D of Figure \ref{denpdf}) the PDF remains very
similar, indicating that the model has reached a quasi-steady state.
At late times, gas is found at both lower and higher densities than
was previously observed before the development of the MRI.  Thus, the
magnetized turbulence induces both strong compressions and significant
rarefactions.  Compared to the maximum ($\rho_{\rm{max}}$) and minimum
($\rho_{\rm{min}}$) densities before the onset of turbulence,
$\rho_{\rm{max}}$ increases by an order of magnitude and
$\rho_{\rm{min}}$ decreases by a factor of about 3.  The fraction of
gas in the intermediate density regime is a factor 2 -- 3 larger after
the full development of MRI compared to early on. The proportion of
thermally-unstable gas is never greater than 20\% of the whole (for
this set of parameters), but exceeds the proportion of
thermally-stable warm gas during the turbulent stages of evolution.

\begin{figure}
\plotone{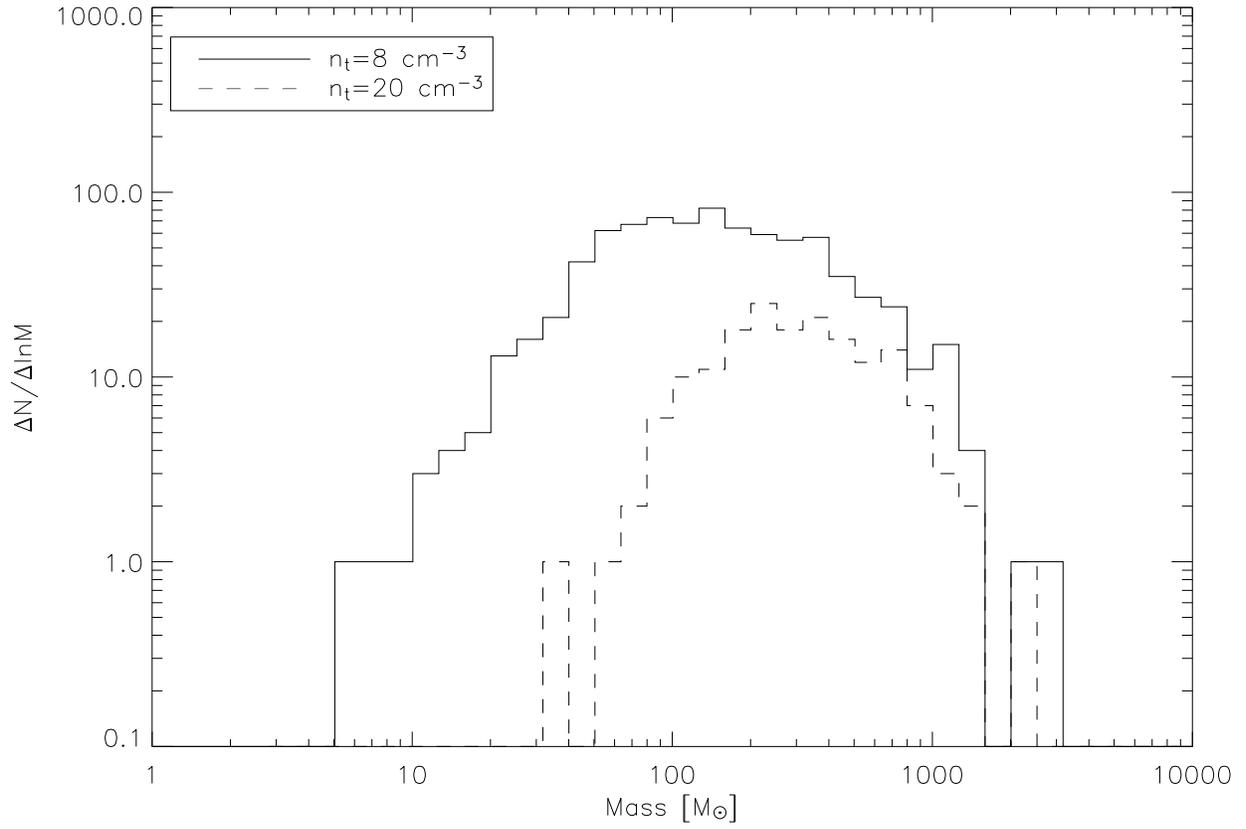}
\caption{Mass spectrum of clumps from snapshot at $t=6.5$ orbits, with threshold densities of 8 and 20 $\cmt$.
\label{mass_dn}}
\end{figure}

To investigate properties of individual condensations in our model, we
use an algorithm similar to that of CLUMPFIND \citep{wil94}.  The
algorithm was developed and applied by \citet{gam03} to identify
clumps in simulations of turbulent molecular clouds.  Briefly, the
algorithm first finds all local maximum values of density in the
computational volume.  All grid cells with a density higher than a
chosen threshold density, $n_t$, are assigned to the nearest local
maximum. This set of continuous zones defines a clump.  The only other
parameter needed is a smoothing length, applied to the initial density
data cube (see Gammie et al. 2003); we set this to 1.5 grid zones.  In
Figure \ref{mass_dn} we show the clump mass spectrum for two different
choices of threshold density, $n_t=8$ and 20 $\rm{cm^{-3}}$.  This
mass spectrum is computed at $t=6.5$ orbits.  Mass spectra from other
late times are similar. With $n_t=8 \ \rm{cm^{-3}}$, 812 clumps were
found, with a minimum clump mass of $5.6 \ M_{\odot}$, and a maximum
mass of $2800 \ M_{\odot}$.  For reference, the total mass in the
simulation is $2.51 \times 10^5 \ M_{\odot}$.  Increasing the critical
density to $n_t=20 \ \rm{cm^{-3}}$, we find 168 clumps, with a minimum
mass of $35 \ M_{\odot}$, and a maximum mass of $2200 \ M_{\odot}$.
For both cases, the peak of the mass spectrum is in the range $100-300
\ M_{\odot}$; the peak increases slightly for larger $n_t$.

\begin{figure}
\plotone{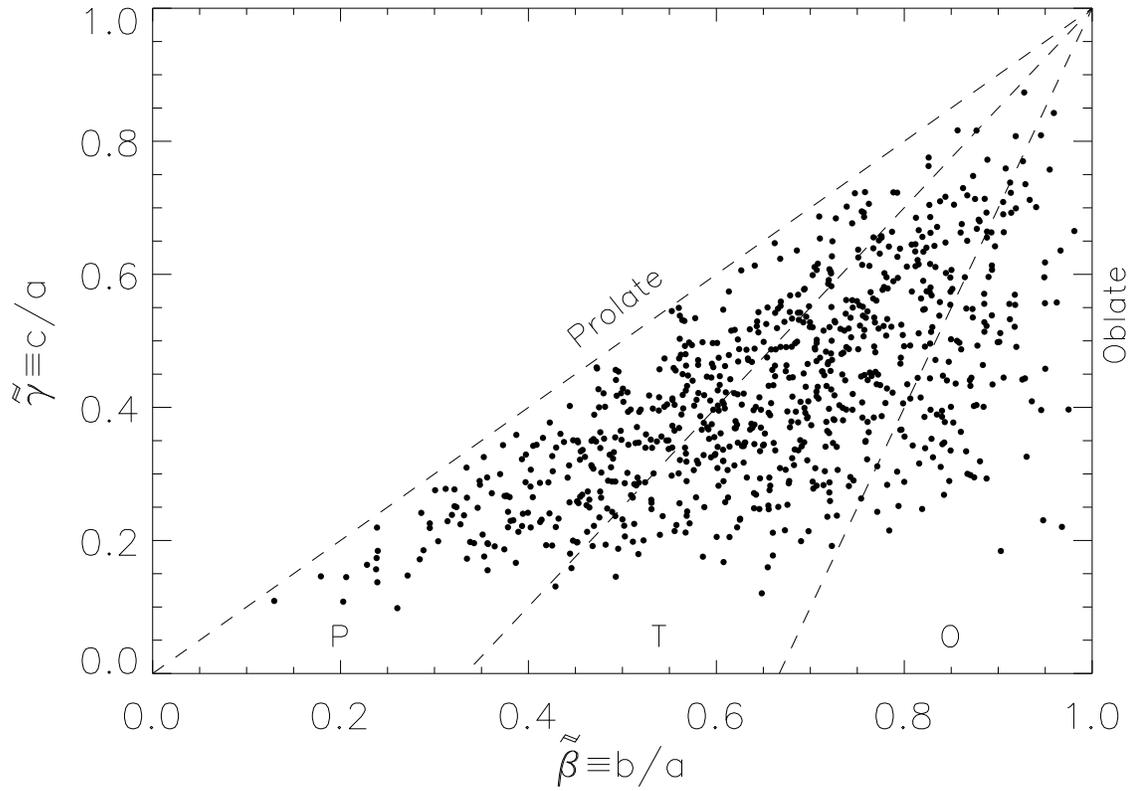}
\caption{Axis ratios for clumps at $t=6.5$ orbits, taking $n_t = 8 \ \cmt$.
\label{shape}}
\end{figure}

To describe their shapes, we compute diagonalized moment of inertia
tensors for each clump, following \citet{gam03}.  Figure \ref{shape}
plots the ratios, for each clump, of the smallest (c) and intermediate
(b) axes to the largest (a) axis.  Prolate-shaped clumps lie near the
diagonal line, oblate clumps lie near the right side vertical axis,
and triaxial clumps lie in the center.  Using two dotted lines to
demarcate these groups, we find 38 \% of the clumps are prolate, 49 \%
are triaxial, and 14\% are oblate. Although clumps are certainly not
round, typical minimum to maximum axis ratios are about 2:1.
``Filaments'', with c/a=0.1 are common, however, and these elongated
structures are easy to pick out in Figure \ref{vol2}.

\subsection{Pressure and Temperature Structure}
\begin{figure}
\plotone{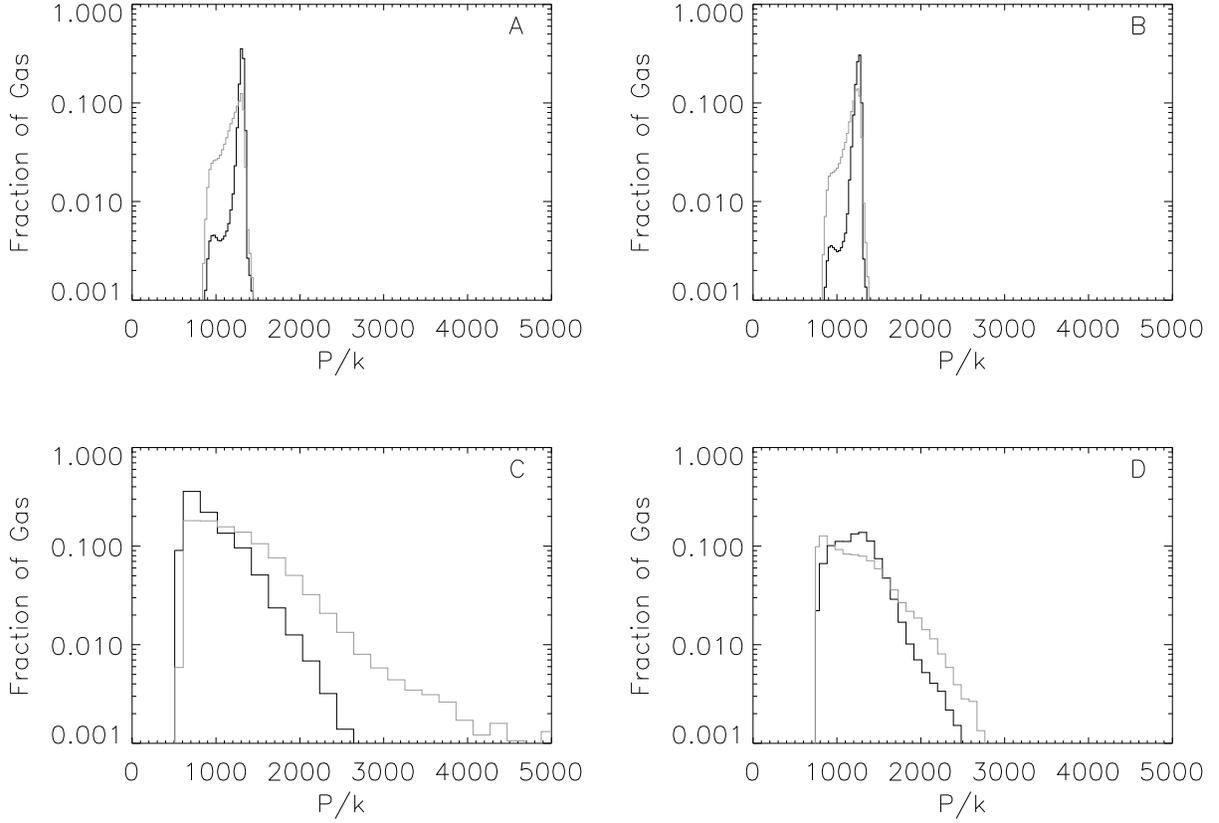}
\caption{Volume (black line) and mass (grey line) PDFs of pressure for the standard run  at $t=1, 2.5, 5.0,$ and $9$ orbits (panels A-D, respectively).  Early in the simulation most of the gas is found in the range of P/k=900-1300 $\rm{K \ cm^{-3}}$.  The MRI has drastically altered the pressure distribution in Panels C and D. 
\label{presspdf}}
\end{figure}

The pressure PDFs at $t=1,2.5,5,$ and 9 orbits are presented in Figure
\ref{presspdf}.  At $t=1$, most of the gas falls within a narrow range
of pressures, P/k=900-1300 $\rm{K \ cm^{-3}}$.  This is lower than
P/k=2000 $\rm{K \ cm^{-3}}$ in the initial conditions, due to
systematic cooling in the thermally unstable stage of evolution.  The
pressure PDF changes little from $t=1$ to $t= 2.5$ orbits, shown in panels A
and B of Figure \ref{presspdf}.  With the development of MRI, however,
gas is driven to both higher and lower pressures, as can be seen in
Panels C and D, at $t=5$ to $t=9$ orbits.  The mean volume-weighted pressure
at the end of the simulation is slightly lower than that after TI has
developed, about P/k=1200 $\rm{K \ cm^{-3}}$.  The pressures in the
cold and warm phases are approximately equal in the latter half of the
simulation, while the pressure in the intermediate phase is slightly
higher, about P/k=1300 $\rm{K \ cm^{-3}}$.  The dispersion in pressure
early in the simulation is about $\delta \rm{P}/{\rm k} \sim 60 \ \rm{K \ 
  cm^{-3}}$, while late in the simulation this increases to as much as
$\delta \rm{P/k} \sim 400 \rm{K \ cm^{-3}}$.

\begin{figure}
\plotone{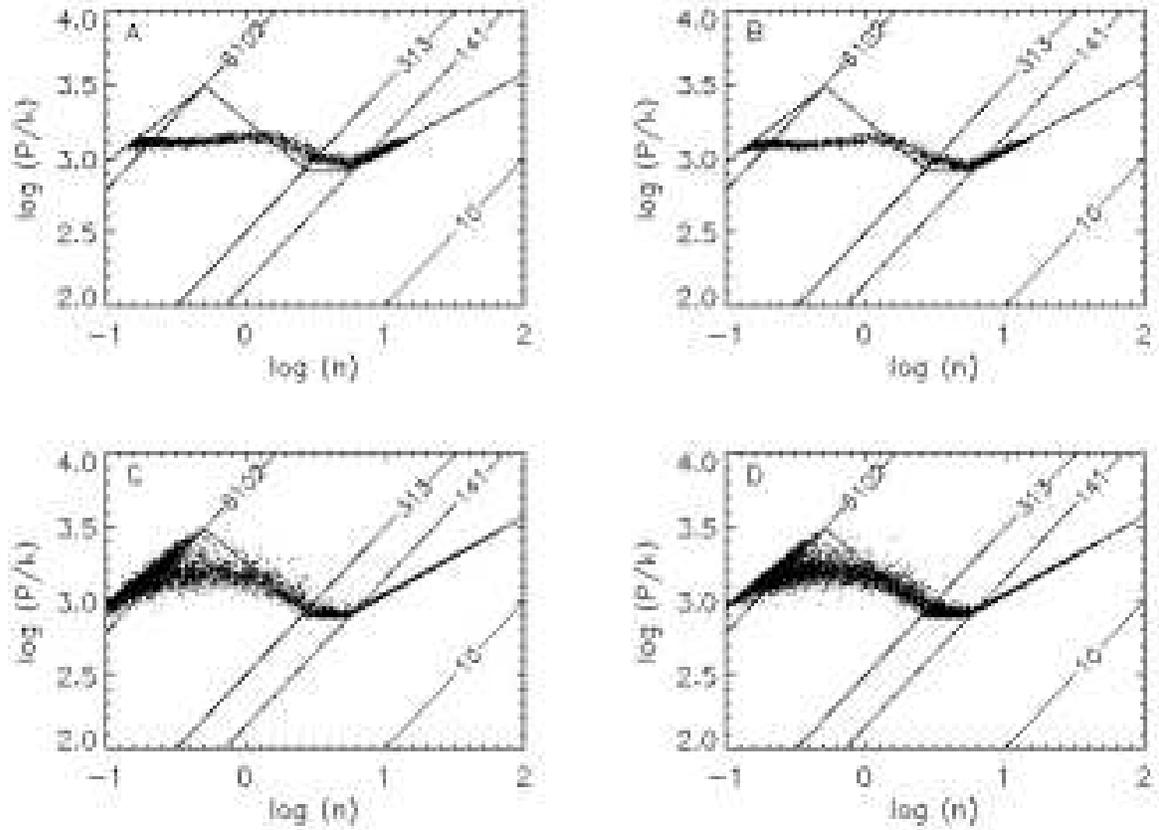}
\caption{Scatter plot of $n$ and $P/k$ for the standard run, at $t=1, 2.5, 5.0,$ and $9$ orbits (panels A-D, respectively).  The equilibrium cooling curve is plotted for comparison, along with temperature contours corresponding to the transitions between the warm, unstable, and cold phases of gas. 
\label{coolcurve}}
\end{figure}

In Figure \ref{coolcurve} we show scatter plots of pressure against
density overlayed on our model cooling curve at $t=1,2.5,5,$ and 9
orbits.  We also show contours of constant temperature to indicate the
transitions between different phases of gas.  Only a fraction of the
zones are included because of the large number of cells contained in
our 3D simulations.  Early in the simulation (Panels A and B), the gas
is close to pressure equilibrium, although high density gas lies
closer to the thermal equilibrium curve.  Later in the simulation
(Panels C and D), strong interactions between clouds can drive gas far
from pressure equilibrium.  At low densities where the cooling
time scale is longer than the dynamical time, gas can be found at
pressures as high as P/k=3200 $\rm{K \ cm^{-3}}$ and as low as 800
$\rm{K \ cm^{-3}}$, a range of a factor four.  Much of the low-density
gas is not in thermal equilibrium.  In high density regions there is
also a wide range of pressures observed (P/k=800-4000 $\rm{K \ 
  cm^{-3}}$), but because the cooling time is very short ($\sim 10^4$
yr) this gas maintains thermal equilibrium.  At early times,
distributions of density and pressure are quite similar to the
corresponding results from our 2D models (Paper I) after the nonlinear
development of TI.  At late times, however, these 3D turbulent models
show much broader pressure distributions than our 2D models.  Overall,
the mean pressure averaged over orbits 6-10 is 1206 $\rm{K \ 
  cm^{-3}}$.  By phase the mean pressure is $\rm{P/k} =$ 1187,
1324, and 1195 $\rm{K \ cm^{-3}}$ in the warm, intermediate and cold
phases.

\begin{figure}
\plotone{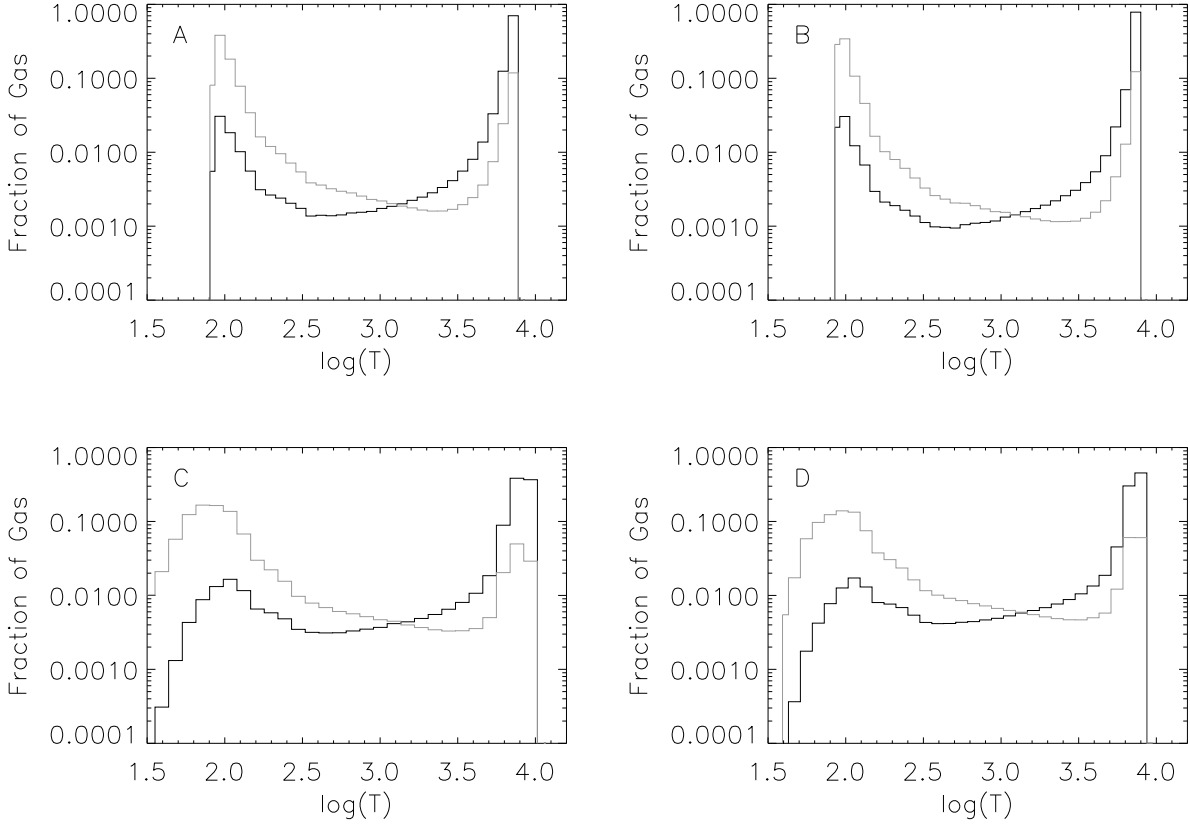}
\caption{Volume (black line) and mass (grey line) temperature PDFs for the standard run  at $t=1, 2.5, 5.0,$ and $9$ orbits (panels A-D).  As for the density PDF, the temperature PDF changes little between Panels A and B.  In Panels C and D the development of the MRI forces gas to higher density, which results in the lower temperatures seen in Panels C and D. In addition, the ranges of temperatures for the warm and cold gas peaks also increase.
\label{temppdf}}
\end{figure}

Also of interest are the temperature PDFs, shown in Figure
\ref{temppdf} at the same times as in Figure \ref{denpdf}. In Panels C
and D, the fraction of gas in the intermediate temperature phase has
increased, and gas is also found at colder temperatures than are
present earlier in the simulation.  The minimum temperature is 80K,
and respectively 60\% and 68\% of the gas mass is found between
80-120K at $t=1$ and $2.5$ orbits.  At $t=5$ and 9 orbits, on the
other hand, respectively 30\% and 18\% of the gas is found at
temperatures below 80K, while respectively another 32\% and 31\% of
gas is at T=80-120 K. The range of temperatures in which the majority
of cold gas is found increases by about a factor of two. The upper
limit on temperature increases slightly throughout the run, but in
addition, the dispersion of temperatures in the warm medium increases.
At early times, $\sim$80\% of the warm gas is in the range
T=6600-8600K, whereas at late times, 80\% is evenly distributed over
twice as large a spread in temperatures.

\begin{figure}
\plotone{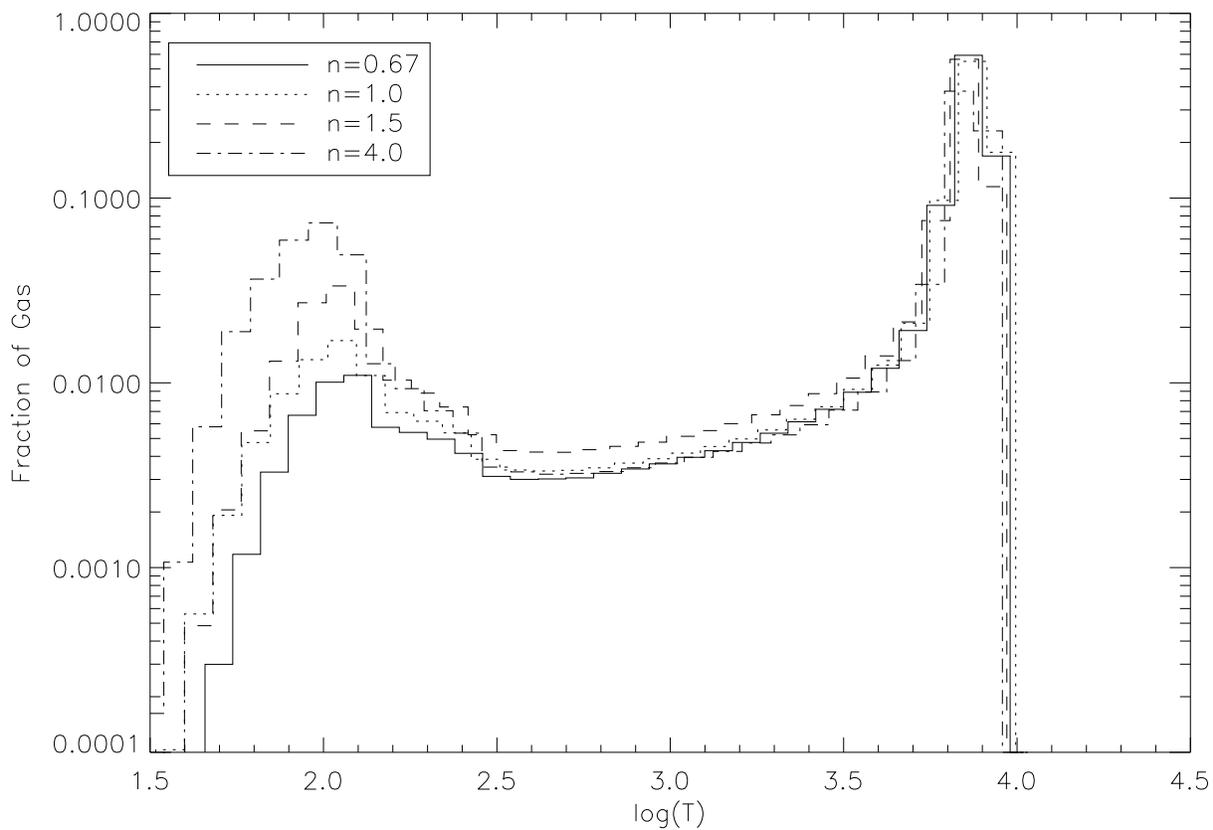}
\caption{Comparison of volume-weighted temperature PDFs for runs with $n=0.67,1.0,1.5,$ and 4.0 $\cmt$, as indicated.  The PDFs are averaged over $t=6.0-6.5$ orbits. The temperature structure is essentially the same, but as the average density is decreased, the cold gas occupies a smaller total volume, lowering the PDF at low temperature.  
\label{tempcomp}}
\end{figure}

\begin{figure}
\plotone{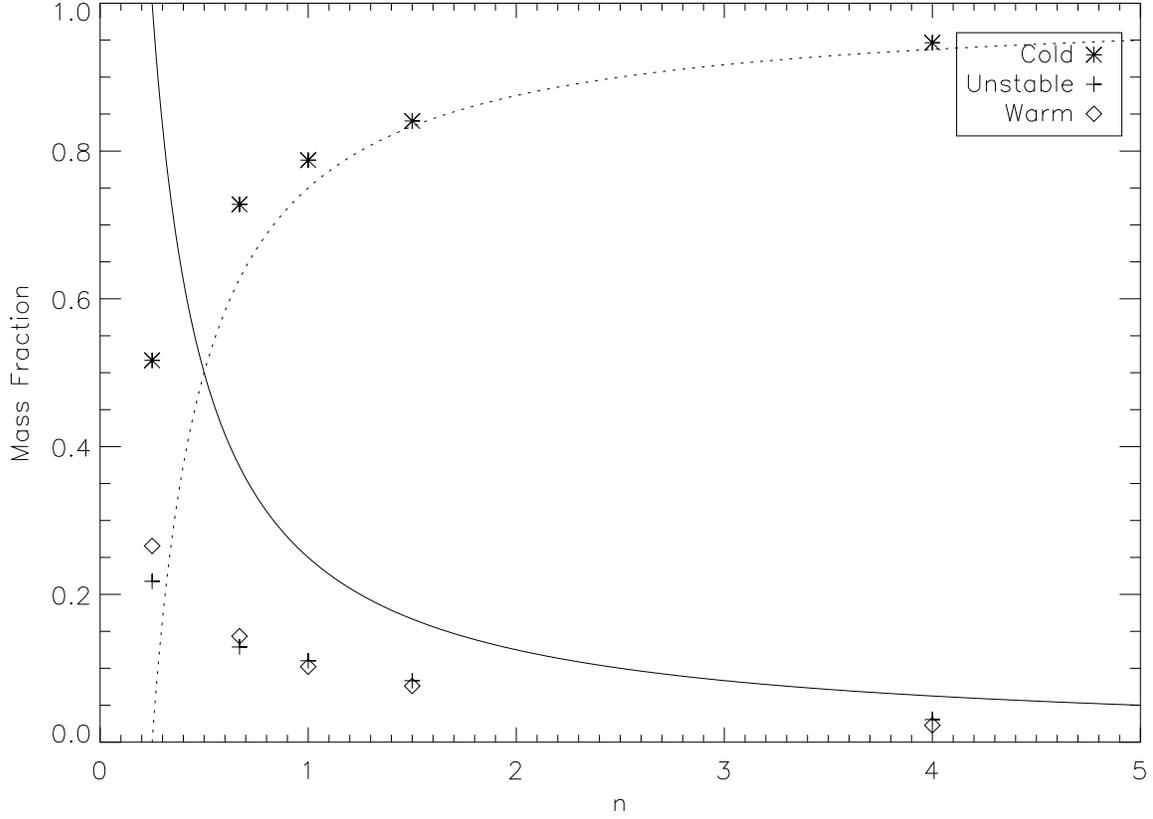}
\caption{Mass fractions for the cold, unstable, and warm phases as a function of mean simulation density, $\bar{n}$. The dotted and solid lines represent the theoretical mass fractions of the cold and warm phases, respectively, for a perfect two-phase medium in pressure equilibrium, assuming the density of the warm medium is $n_w=0.25 \ {\rm cm^{-3}}$, which is typical for our simulations.
\label{massfrac}}
\end{figure}

Figure \ref{tempcomp} compares the volume-weighted temperature PDFs of
four runs of different mean density.  These four runs have average
densities of $\bar{n}=4.0, 1.5, 1.0$, and $0.67 \ \rm{cm^{-3}}$ and,
as we shall discuss in \S \ref{kinetics}, the mean velocity dispersion
increases by an order of magnitude from the highest to lowest mean
density models.  The PDFs in Figure \ref{tempcomp} represent averages
from 6.0-6.5 orbits.  At intermediate and high temperatures, the PDFs
for these runs are quite similar.  Most of the warm gas is at
T=6000-8000 K, with $\rm{T_{max}} \approx$ 10000 K.  Most of the cold
phase is at temperatures near 100 K, possibly showing a slight trend
towards higher mean temperature as $\bar{n}$ is decreased.  Overall
there is less gas at lower temperature when $\bar{n}$ is reduced,
because the total mass available for cold clouds is lower.  In
addition to having similar warm and cold gas temperatures, the models
with various $\bar{n}$ are similar in that the fractions of gas in the
intermediate- and warm-temperature regimes are always quite close.
These results are illustrated in Figure \ref{massfrac}, which plots
the mass fractions in the various regimes as a function of $\bar{n}$
(also including the $\bar{n}=0.25$ model).  

Overplotted in Figure \ref{massfrac} are curves indicating the warm
and cold gas mass fractions that a pure two-phase medium would have.
The mass fraction of cold gas in a perfect two-phase medium in thermal
and pressure equilibrium is $f_c=(1-n_w/\bar{n})/(1-n_w/n_c)\approx
1-n_w/\bar{n}$, where $n_c$ is the cold density, $n_w$ is the warm
density, and $\bar n$ is the mean density.  The mass fraction of warm
gas is then $f_w\approx n_w/\bar{n}$.  The density of warm gas in our
simulations is typically $n_w=0.25$, which we use to compute the
theoretical curves in Figure \ref{massfrac}.

The possibility exists that our choice of initial conditions in the
standard run, a uniform medium at the average density, may have some
effect on the amount of gas in the intermediate phase at late times.
Due to TI, initially most of the gas collects into small , dense cold
clouds, and only a small proportion of the gas remains in the
thermally unstable regime.  Later in the simulation, the MRI drives a
larger fraction of gas into the unstable phase.  It is possible that
if we had {\it begun} with a turbulent medium, this fraction would be
even larger, from increased shock heating of moderate density clouds
with larger collision cross sections.  To investigate this, we
initialized a simulation with the same mean density and magnetic field
as our standard run, but evolved it with an isothermal equation of
state.  The sound speed was set so that the initial $P/k$ matches late
time averages from our standard run.  After the isothermal evolution
has proceeded for 10 orbits and reached a saturated turbulent state,
heating and cooling are enabled.  After a quasi-steady state is
reestablished, we measure the mass fractions in the warm,
intermediate, and cold regimes.  The result is respective proportions
of about 11\%, 14\%, and 75\%, which is similar to our results from
standard run.  Thus, we conclude that the long-term thermal history
does not strongly affect the present state of the gas.

\subsection{Turbulent Velocities}
\label{kinetics}
\begin{figure}
\plotone{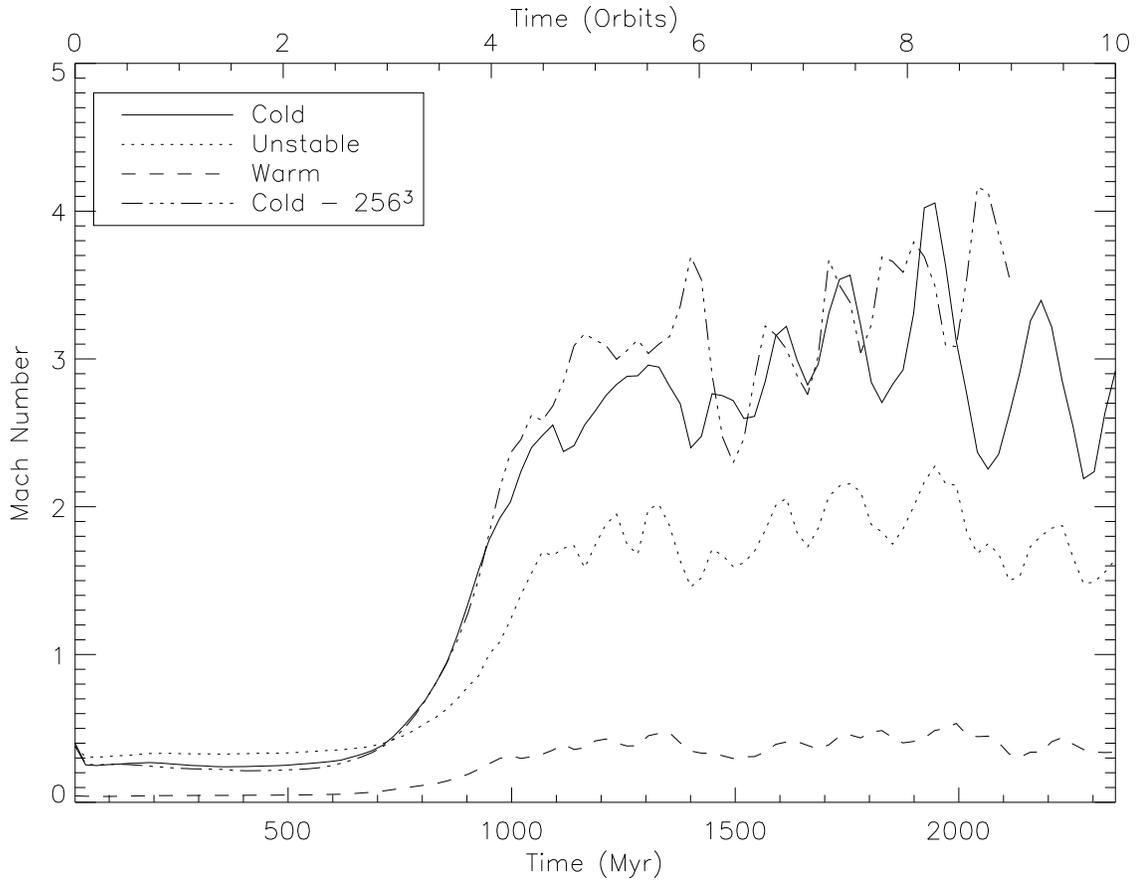}
\caption{Mass weighted Mach number for the three phases of gas in the standard model.  The average Mach number for the second half of the simulation is 0.4, 1.8, and 2.9 for the warm, unstable, and cold phases.  For comparison we also plot the mass weighted Mach number of the cold medium for the high resolution $256^3$ run.
\label{mach}}
\end{figure}

In Figure \ref{mach} we plot the mass-weighted Mach number ${\cal
  M}\equiv \delta v/c_s$ of the gas in each thermal phase (warm,
intermediate, cold) as a function of time for the duration of the
simulation.  We also include, for comparison, the mass-weighted Mach
number of the cold medium for the high resolution run at $256^3$.  The
isothermal sound speed $c_s=(\rm{kT}/\mu)^{1/2}$ is computed
individually for all grid zones, and the galactic shear is subtracted
from the azimuthal ($v_y$) velocity before computing
$\delta v^2=v_x^2+(\delta v_y)^2+v_z^2$.  Initially, motions in all three
phases of the gas are subsonic, ${\cal M} < 0.3$, and remain so until
the MRI begins to develop at about 800 Myr ($\sim 3$ orbits).  Once
the MRI saturates (at $t \sim 5$ orbits), the typical Mach numbers of
the warm, intermediate and cold phases of the gas are 0.4, 1.8, and
2.9.  The peak value of ${\cal M}$ for the cold phase is about 3.2.
The mean late time velocity dispersion for all three phases of the gas
is similar, approximately $2.7 \ {\rm km \ s^{-1}}$.  At late times,
the individual velocity dispersions in the radial, azimuthal, and
vertical directions are 1.9, 1.7, and 0.7 \ ${\rm km \ s^{-1}}$,
respectively.

\begin{figure}
\plotone{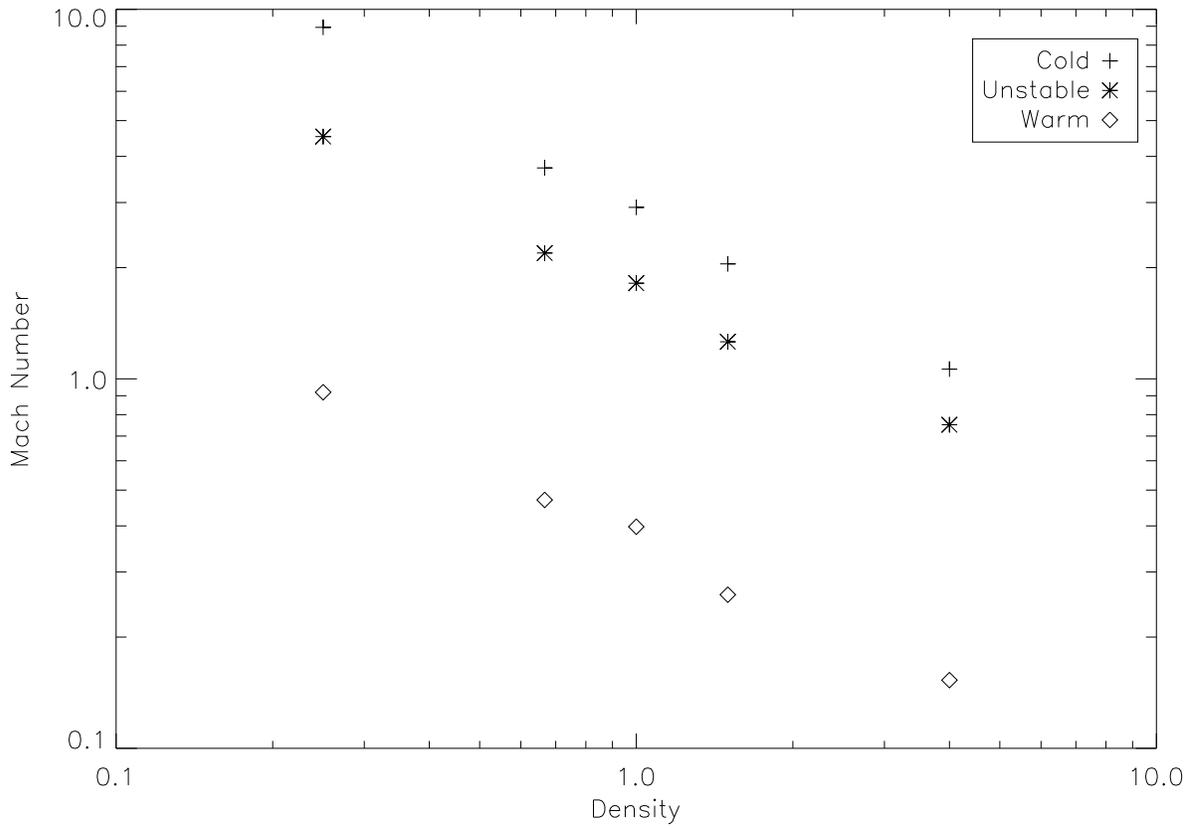}
\caption{Mach number, separated by phase, plotted against the average density in the box for five different simulations. Mean densities of models are $\bar{n}=0.25,0.67,1.0,1.5$ and $4.0 \ {\rm cm^{-3}}$.  Linear fits to the results give power law slopes of -0.67, -0.68 and -0.77 for the warm, unstable, and cold phases, respectively.  
\label{machplot}}
\end{figure}

To explore the dependence of saturated state turbulence on system
parameters, for our five simulations of varying mean density $\bar{n}$
we have computed the average Mach number over $t=5-10$ orbits. We plot
the results, separating the three thermal phases, as a function of
$\bar{n}$ in Figure \ref{machplot}.  The relationships between ${\cal
  M}$ and $\bar n$ clearly follow power laws.  The slopes for the
warm, intermediate and cold phases are $d\ln {\cal M}/d\ln \bar n =
-0.67, -0.68$ and -0.77.  Since the cold component dominates the mass,
this implies $(\delta v) \propto (\bar n)^{-0.77}$ overall.  For our
$\beta=1000$ model at $\bar{n}=1 \ {\rm cm^{-3}}$, the saturated state
Mach numbers are 0.3, 1.1, and 1.6 for the warm, intermediate, and
cold phases.  Our results are thus consistent with general findings
from previous MRI simulations that saturated-state turbulent
amplitudes increase with increasing mean Alfv\'{e}n speed.  The
detailed scalings, however, show interesting differences, which we
shall discuss in \S \ref{discuss}.

We have found that the turbulence is quite insensitive to
particularities of structure in initial conditions.  Thus, our model
which began with a two phase ``cloudy'' medium, with the same initial
mean density as our standard run, saturates with nearly the same
velocity amplitude as the standard run. The initially-isothermal run
which was restarted with cooling also yielded similar results to the
standard run, with Mach numbers of 0.4, 1.7, and 2.8 for the warm,
intermediate and cold phases. The saturated state of the isothermal
simulation itself has a Mach number of 1.4, corresponding to mean
velocity dispersion $4.0 \ \rm{km \ s^{-1}}$, somewhat larger than for
our cooling models at this fiducial mean density.  Differences between
isothermal and multiphase models are likely to depend on $\bar{n}$,
however.

\begin{figure}
\plotone{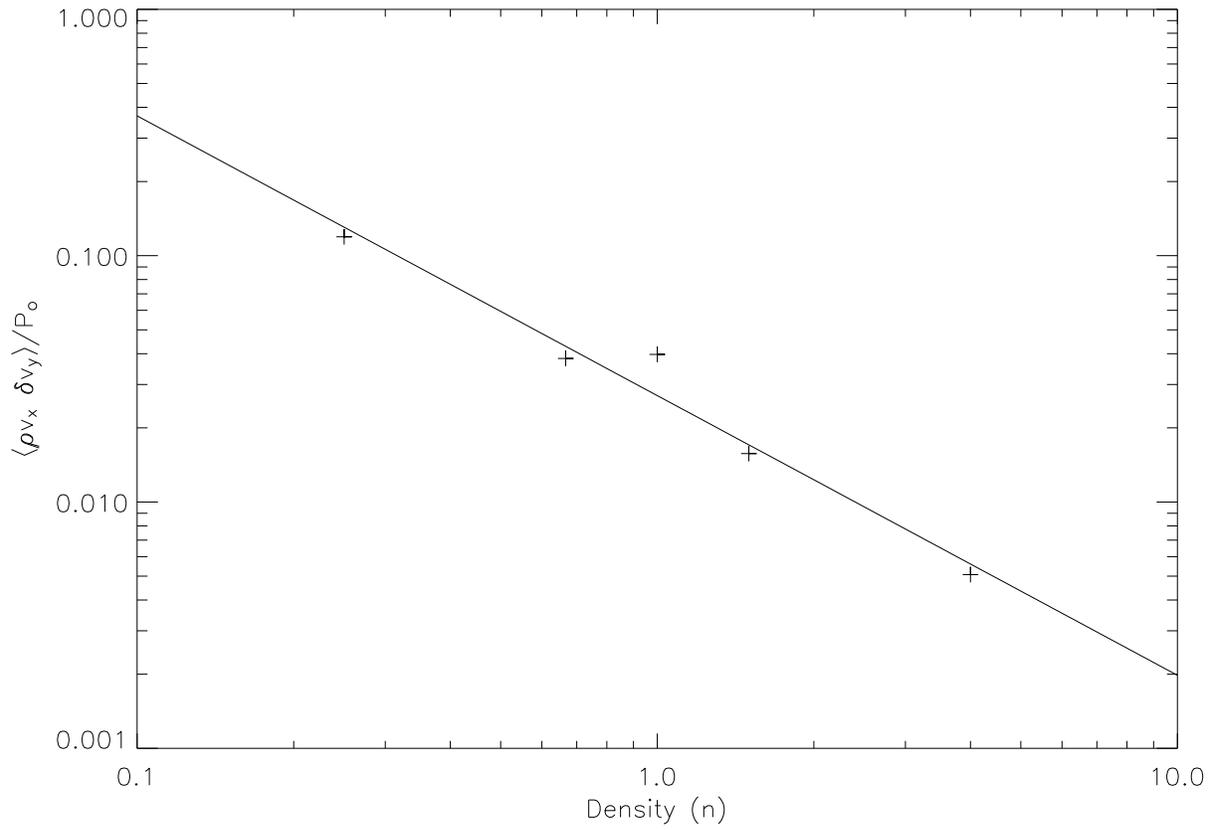}
\caption{Reynolds stress plotted against the mean density for five different simulations with $\bar n=0.25,0.67,1.0,1.5$ and $4.0 \ {\rm cm^{-3}}$.  A fit gives a power law slope of -1.1.
\label{reynoldsplot}}
\end{figure}

The average Reynolds stress, $\langle\rho v_x \delta v_y\rangle/P_0$,
from $t=5-10$ orbits is plotted against the mean density for $\bar n =
4.0, 1.5, 1.0,0.67,$ and 0.25 $\cmt$ in Figure \ref{reynoldsplot}.
The relationship again follows a power law, with a slope of -1.1.  

The velocity power spectra are generally consistent with
previous simulations of the MRI \citep{haw95a, kim03}.  The largest
scales dominate the simulation, generally following a Kolmogorov-like
spectrum, $\sim k^{-11/3}$.  Our quoted values for the velocity
dispersions therefore correspond to the largest scales in the
simulations.  On smaller scales, such as an individual cloud, the
velocity dispersion would be smaller.  We have tested the relation
between linewidth and size directly, using the ``ROC'' analysis
approach described in \citet{ost01}.  Both for the cold component
alone, and for the whole medium, we find that the velocity dispersion
increases with the size of clouds, or sub-boxes of the computational
volume.

\subsection{Magnetic Fields}

\begin{figure}
\plotone{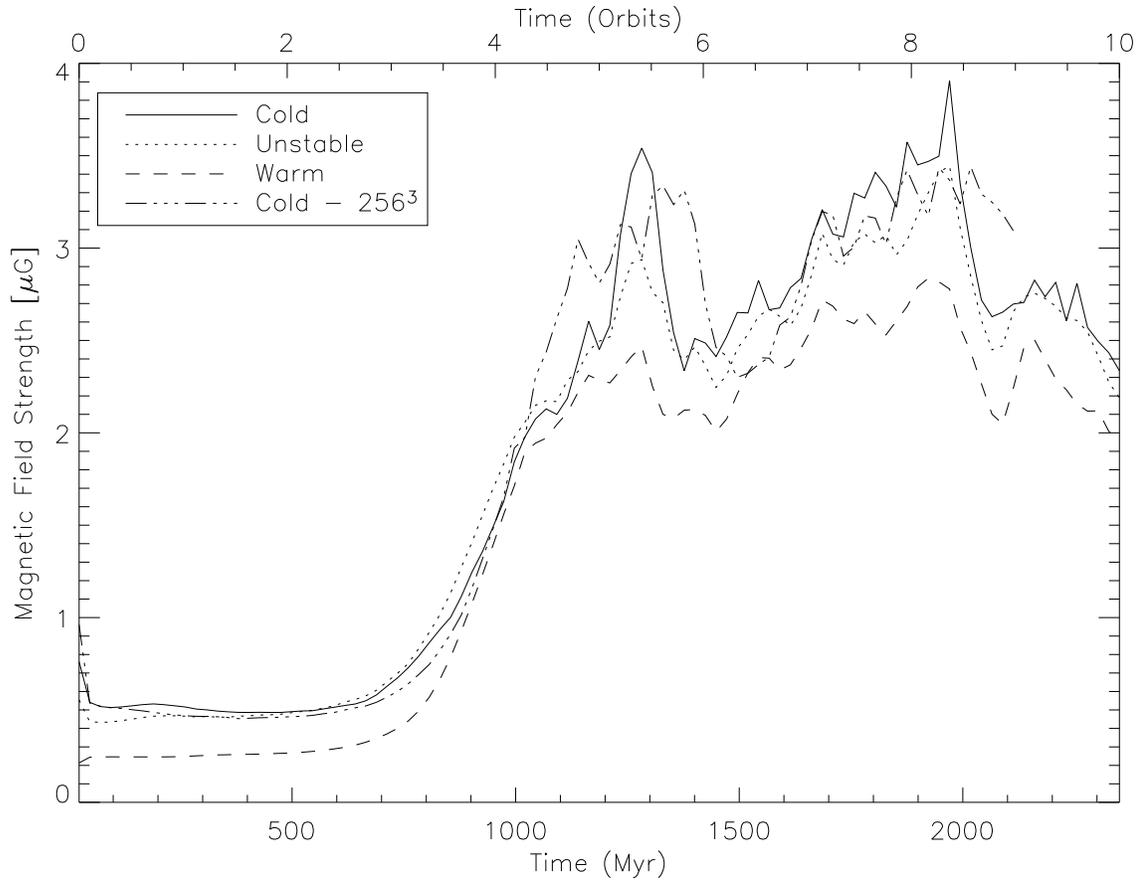}
\caption{Magnetic field strength plotted against time for standard run (with initial $\beta = 100$, i.e. ${\rm (B_z)_{init}}= 0.26 \ \mu$G). The typical saturated state field strength is $2-3 \ \mu$G, with little difference between the three phases.  For comparison we also plot the field strength in the cold medium for the high resolution $256^3$ run.
\label{bfield}}
\end{figure}

Similarly to the (random) kinetic energy, the magnetic energy
increases as the MRI develops.  In Figure \ref{bfield} we plot the
magnetic field strength as a function of time for each of the three
phases of gas.  In the initial conditions, $ B=B_z=0.26 \ \mu$G.
After TI develops, the field strength is $0.25 \ \mu$G for the warm
phase, and about $0.5 \ \mu$G for the (denser) unstable and cold
phases.  As the MRI develops, after $t=5$ orbits, the field strength
grows to range over $2-3 \ \mu$G for all three phases, reaching as
high as $4.1 \ \mu$G in the cold phase.  The late time component
magnetic field strengths, $\langle B_x^2 \rangle ^{1/2}, \langle B_y^2
\rangle ^{1/2},$ and $ \langle B_z^2 \rangle ^{1/2}$ are 1.3, 1.9, and
0.51 $\mu$G, averaged over $t=6-10$ orbits.  Thus, the MRI enhances
the magnetic field by an order of magnitude over its initial value.
We note that if overdense clouds were to form by isotropic contraction
of the ambient medium, then one would expect $\langle
B^2\rangle^{1/2}\propto \rho^{2/3}$.  With a cold medium density two
orders of magnitude larger than that of the warm medium, the
respective mean field strengths would differ by a factor 20.  Since
this is not the case, condensation evidently proceeds preferentially
along field lines.

\begin{figure}
\plotone{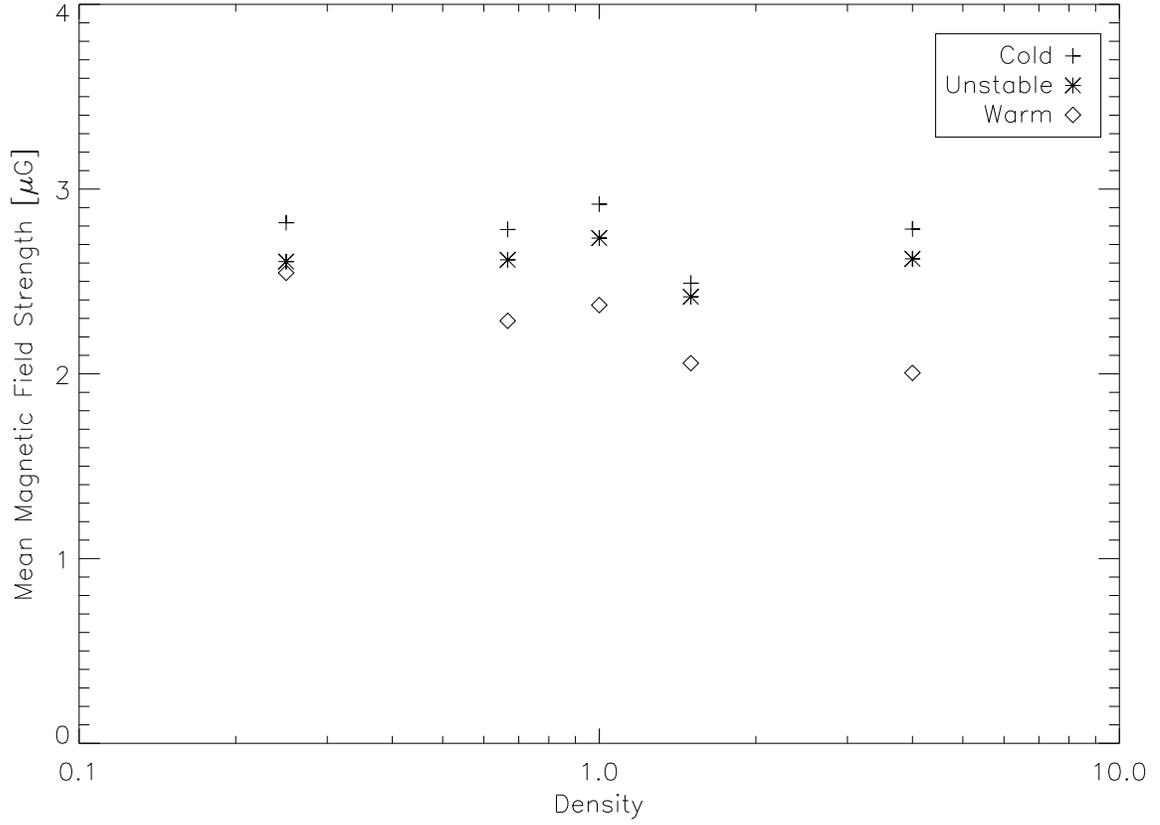}
\caption{Late time magnetic field strength as a function of density for five runs with $\bar{n}=0.25,0.67,1.0,1.5$ and $4.0 \ {\rm cm^{-3}}$.  
\label{magplot}}
\end{figure}

To explore dependence on mean properties, in Figure \ref{magplot} the
late time magnetic field strength, averaged over five orbits, is
plotted against the mean density in the box for five simulations with
$\bar n = 4.0, 1.5, 1.0,$ and 0.67 ${\rm cm^{-3}}$.  Unlike the
turbulent velocity dispersions, the $B$ field strength does not show
any significant trend with $\bar{n}$, saturating between 2 and 3
$\mu$G.  The field strength also does not differ significantly between
the cold, intermediate, and warm phases for any of the models.  As a
marginal effect, the field strength in the warm medium decreases as
$\bar{n}$ increases.

\begin{figure}
\plotone{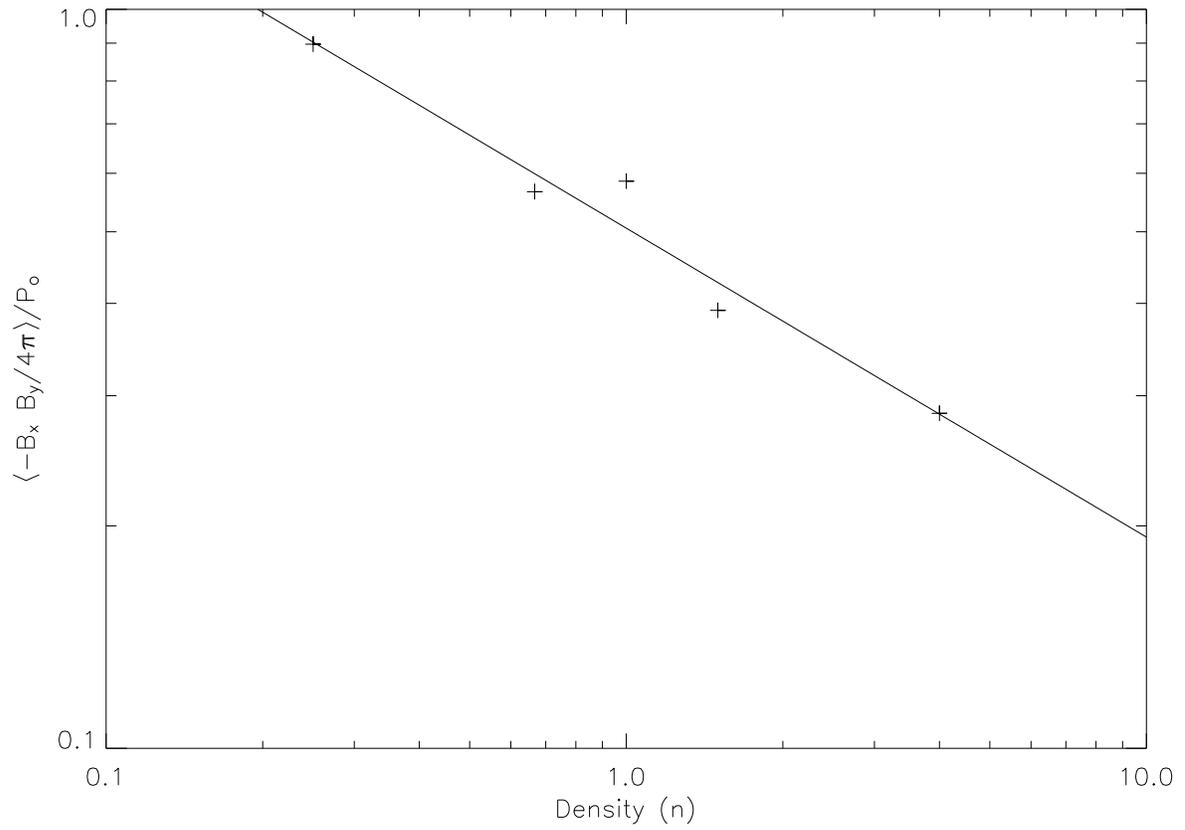}
\caption{Maxwell stress as a function of mean density for five runs with $\bar{n}=0.25,0.67,1.0,1.5$ and $4.0 \ {\rm cm^{-3}}$. A fit gives a power-law slope of -0.42.
\label{maxplot}}
\end{figure}

Unlike the magnetic energy density, the Maxwell stress does
show dependence on $\bar{n}$. This stress, $\langle -B_xB_y/4\pi
\rangle /P_0$, is averaged over $t=5-10$ orbits and plotted against
the mean density for five simulations with $\bar n = 4.0, 1.5, 1.0,0.67$
and 0.25 $\cmt$ in Figure \ref{maxplot}.  For the data shown, a power law fit
yields slope -0.42.  Previous single-phase MRI simulations show
somewhat different scalings of Maxwell stresses and magnetic energies,
as we shall discuss in \S\ref{discuss}.

The power spectra of the magnetic field, like the velocity power
spectra, is consistent with previous simulations of the MRI
\citep{haw95a, kim03}, dominated by the largest scales and generally
following a Kolmogorov-like spectrum.

\subsection{Energetics}

Tracking the changes in various energies is key to understanding the
interrelationships between dynamics and thermodynamics in turbulent
flows.  For the models we have performed, the ultimate energy source
is the shear flow, which drives the MRI.  In turn, turbulent
dissipation can convert kinetic and magnetic energy to thermal energy,
which can subsequently be lost to radiation.  More formally, following
HGB1, we consider the average over the box of the total energy per unit
volume,
\begin{equation}
\langle {\cal H} \rangle = \left\langle \rho\left(\frac{1}{2}v^2 + \frac{{\cal E}}{\rho}-\frac{q \Omega^2x^2}{2}\right)+\frac{B^2}{8\pi}\right\rangle.
\end{equation}
Changes to this energy can occur due to losses or gains from
radiation, and from fluxes through and stresses on the surface of the
computational volume.  With shearing-periodic boundary conditions, the
net rate of change should ideally obey
\begin{equation}
\frac{d}{dt}\langle {\cal H} \rangle = q\Omega \left\langle \rho v_x \delta v_y - \frac{B_xB_y}{4\pi}\right\rangle + \langle -\rho{\cal L} \rangle.
\end{equation}
Thus, if quasi-steady state is reached, we would then expect $d\langle{\cal H}\rangle/dt=0$, and the sum of stresses times $q\Omega$ to equal the cooling rate.
In steady state, from equation (3) the total rate of work done by the combination of
compressions and shocks, $\langle -P\nabla \cdot {\bf v} \rangle +
\langle \left(\frac{\partial {\cal E}}{\partial
    t}\right)_{shocks}\rangle$, plus any other dissipation, should
also be balanced by the net cooling, $\langle \rho {\cal L} \rangle$.

\begin{figure}
\epsscale{0.75}
\plotone{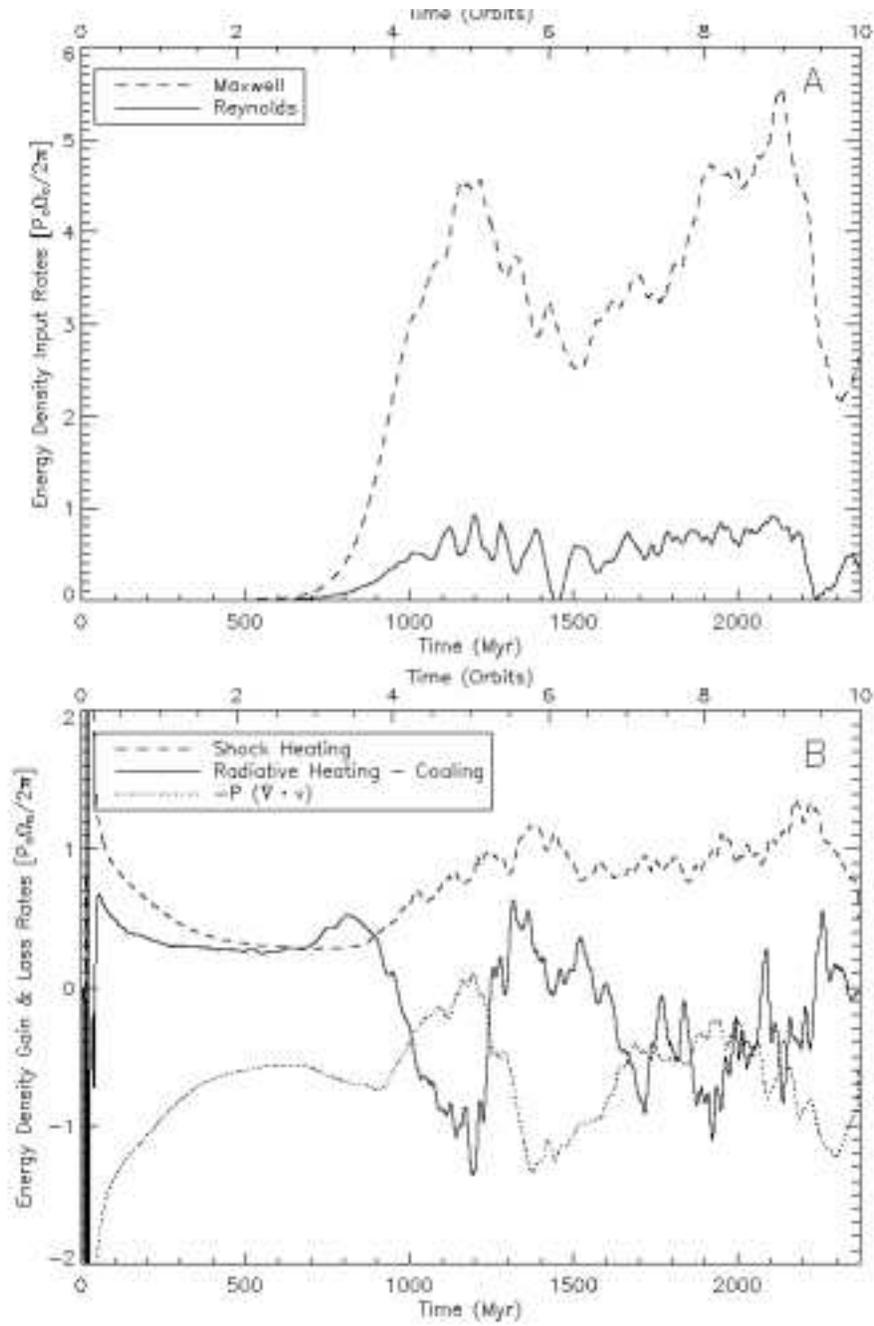}
\caption{In Panel A (upper) we show volume averaged energy input rates from Maxwell and Reynolds stresses plotted against time for a run with standard parameters.  In Panel B (lower) we plot the volume averaged energy density gain and loss rates of shock heating, radiative heating - cooling, and $-P(\nabla \cdot v)$ work, plotted against time, for the same run.
\label{energy1}}
\end{figure}

In the upper panel of Figure \ref{energy1} we plot (for our another
realization of our standard run) the rate of work done by Reynolds and
Maxwell stresses per unit volume.  The late-time volume-averaged
energy inputs from Maxwell and Reynolds stresses are 3.7 and 0.6 (in
units of ${\rm P_0}\Omega/2\pi$).  The Maxwell stress dominates the
Reynolds stress, which is typical in simulations of the MRI.  In the
lower panel of Figure \ref{energy1} we plot the volume averaged shock
heating, radiative heating - cooling rates, $-\langle \rho {\cal
  L}\rangle$, and pressure work, $-\langle P\nabla \cdot v\rangle$, as
a function of time.  The sum of these three terms is approximately
zero during the first few orbits of the simulation.  Later in the
simulation there is typically either net heating or cooling at any
particular time, but the late time averages - individually, 1.0, -0.7,
and -0.3 (in units of ${\rm P_0}\Omega/2\pi$) for shock, radiation,
and pressure terms - sum to zero.  Thus, on average, radiative losses
exceed radiative gains, cooling by rarefactions exceeds heating by
compressions, and together these net loss terms balance gains in
shocks.  The mean energy density typically varies by 10\% during the
latter half of the simulation.

If total energy were perfectly conserved, as noted above, the energy
inputs from Maxwell and Reynolds stresses would be balanced by net
cooling.  The energy source for this radiative cooling would, in turn,
be provided by compressive work and dissipation of turbulence.  The
present simulation, however, in fact captures only part of the
turbulent dissipation -- that in shock heating, as mediated by
artificial viscosity.  In addition, both magnetic and kinetic
turbulent energy are lost at the grid scale.  Oppositely-directed
magnetic fields and shear flows, when advected into a single zone, are
averaged to zero.  Since ZEUS evolves a (non-conservative) internal
energy equation rather than a total energy equation, the associated
energy from those small-scale sheared $\bf{v}$ and $\bf{B}$ fields is
lost. In principle, these dissipation terms could be captured if
explicit resistivity and shear viscosity were included.

%
%

\subsection{Synthetic Line Profiles}
\label{lines}
\begin{figure}
\plotone{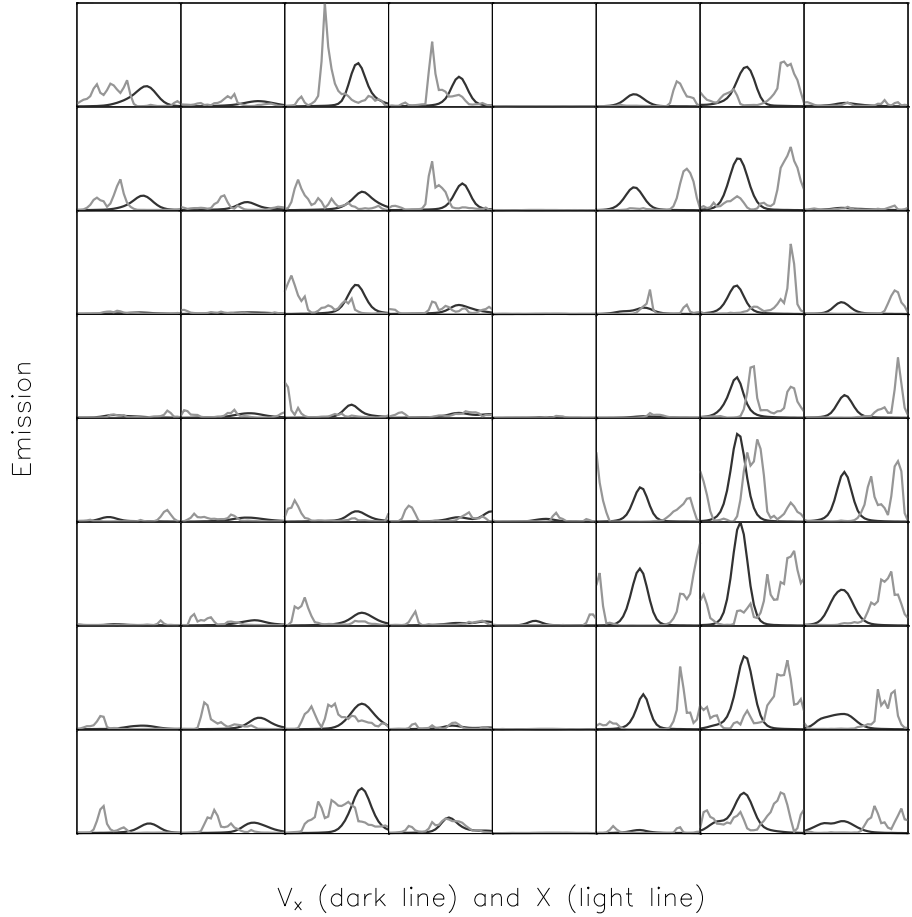}
\caption{Position and velocity profile map in Y-Z plane, for standard run projected along $\hat x$.  Dark lines show integrated emission profiles as a function of $v_x$ along the line-of-sight direction.  Light lines show profiles of emission as a function of $x$ integrated over line-of-sight velocity.
\label{line_x}}
\end{figure}

\begin{figure}
\plotone{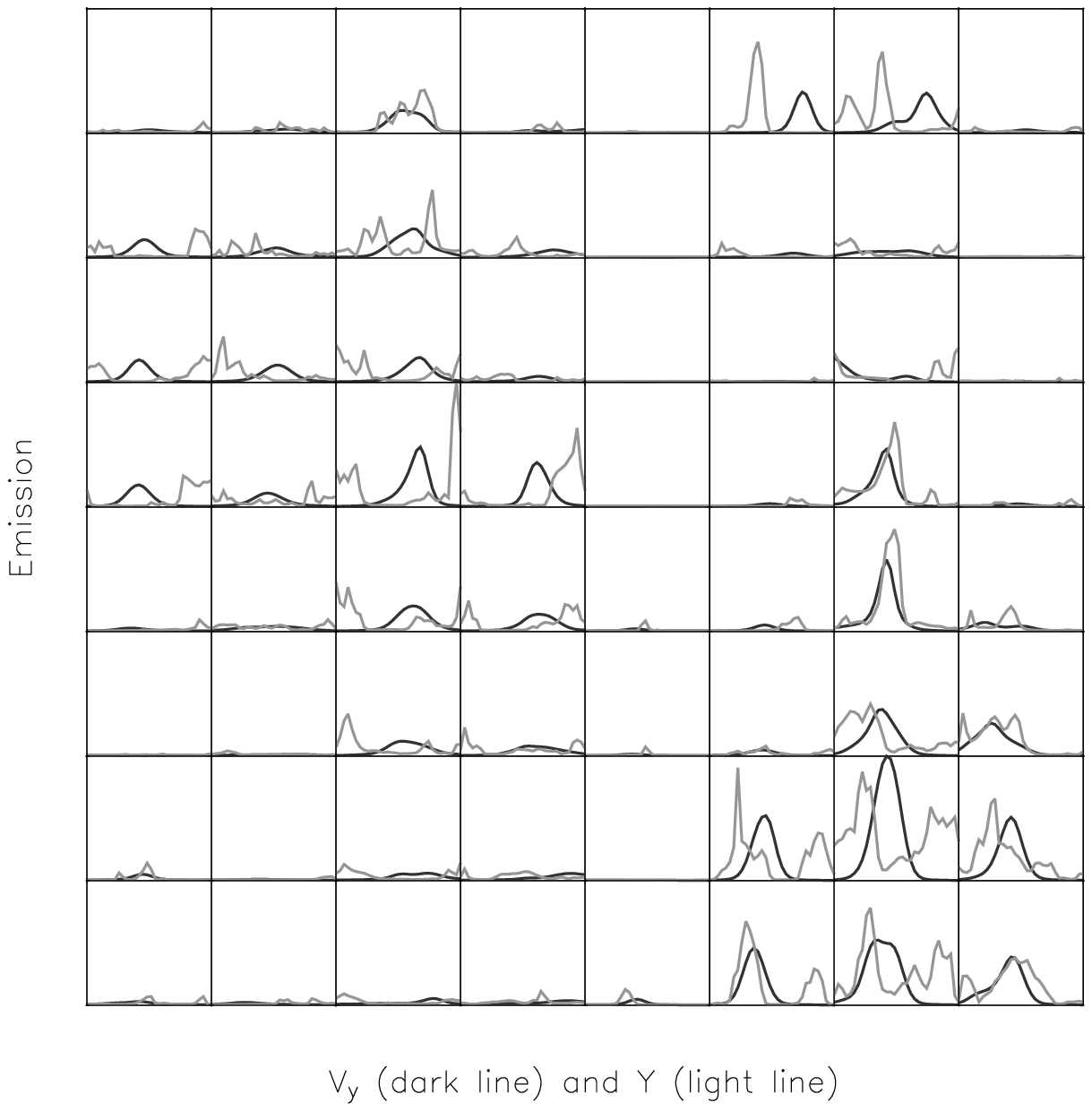}
\caption{Same as Figure \ref{line_x}, for X-Z map projected along $\hat y$.
\label{line_y}}
\end{figure}

\begin{figure}
\plotone{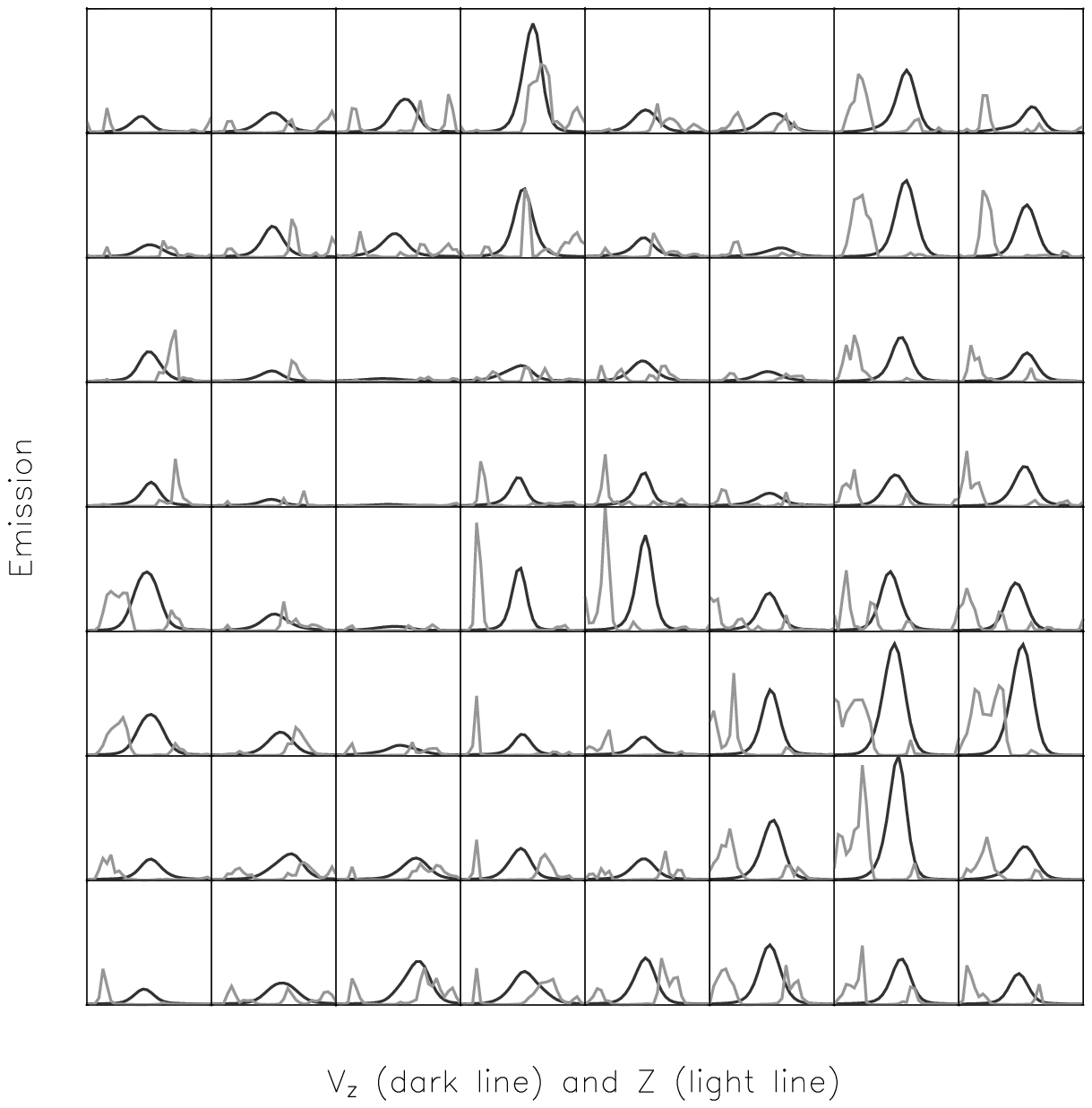}
\caption{Same as Figure \ref{line_x}, for X-Y map projected along $\hat z$.
\label{line_z}}
\end{figure}

Although the present simulations are highly idealized in many ways
(e.g. they are vertically periodic rather than stratified), it is
interesting to explore model properties that bear a close relation to
observables.  The profiles of 21 cm HI absorption directly trace the
density, temperature, and turbulent velocities of the atomic ISM via a
line-of-sight convolution.  Using our simulated ``data,'' we can
generate analogous maps of line profiles projected in any direction
through the computational volume.  Figures \ref{line_x}, \ref{line_y},
and \ref{line_z} show synthetic emission profile maps for our standard
model along the x, y, and z directions.  We also present, paired with
each line-of-sight velocity profile, the corresponding distribution of
total emission with line-of-sight position. Each of the $8 \times 8$
windows on the map represents a volume of $32 \times 32 \times 256 $
zones, integrated over the projected area. For each zone, the
contribution to emission is proportional to the density, with a
Gaussian velocity distribution centered on the flow velocity, and dispersion
$=\sqrt{{\rm kT}/\mu}$.  Strong lines indicate more total mass along a
given line of sight, and weak lines indicate less mass.  Since most of
the mass in our standard model is in the cold phase, a strong line indicates
the presence of cold, dense gas.

%



For Figures \ref{line_x}, \ref{line_y}, and \ref{line_z} the mean line
widths of the velocity profiles are 1.8, 1.9, and 1.7 $\rm{km \ 
  s^{-1}}$.  Without thermal broadening the line widths are reduced to
1.0, 1.3, and 0.7 $\rm{km \ s^{-1}}$, which is consistent with the
time-averaged velocity dispersions in the radial, azimuthal, and
vertical directions reported in \S \ref{kinetics}.  Most of the
velocity profiles are single-peaked, and would likely be interpreted
as arising from one to three emitting components if a standard
Gaussian fitting procedure were applied.  However, our results show
that in many cases several spatially-separated components are
distinctly evident in the line-of-sight mass distributions.  Velocity
profile broadening and skewness statistically give evidence that more
than one component is present, but we find no correlation between
increased spatial coherence and increased velocity profile symmetry in
any given direction.  We do not observe any structure which shows two
distinct lines. The absence of velocity profiles with two distinct
peaks owes in part to the thermal broadening, which smears out smaller
scale features.  However, the primary reason that profiles are
single-peaked is that velocity modes at a large range of wavenumbers
are present in saturated state MRI-driven turbulence.  Since there is
no single dominant wavelength along the line-of-sight, the range of
velocities is smoothly filled.

\begin{figure}
\plotone{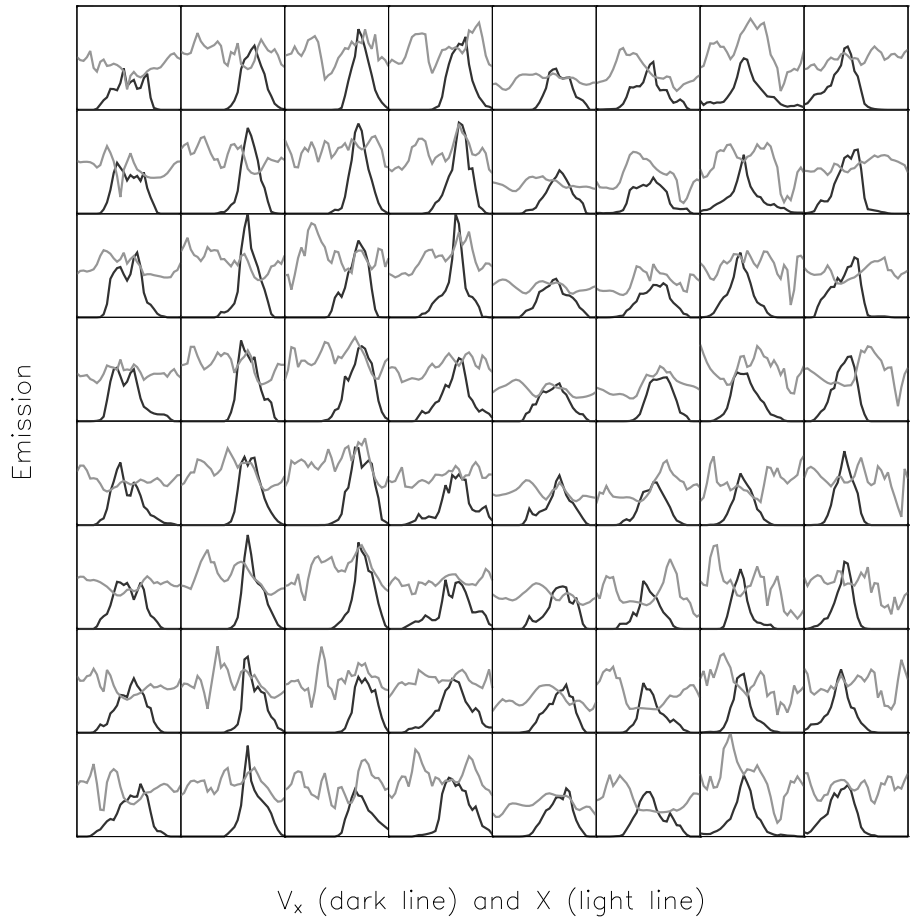}
\caption{Same as Figure \ref{line_x}, for warm gas only, without thermal broadening.
\label{line_x_warm}}
\end{figure}

\begin{figure}
\epsscale{1.}
\plotone{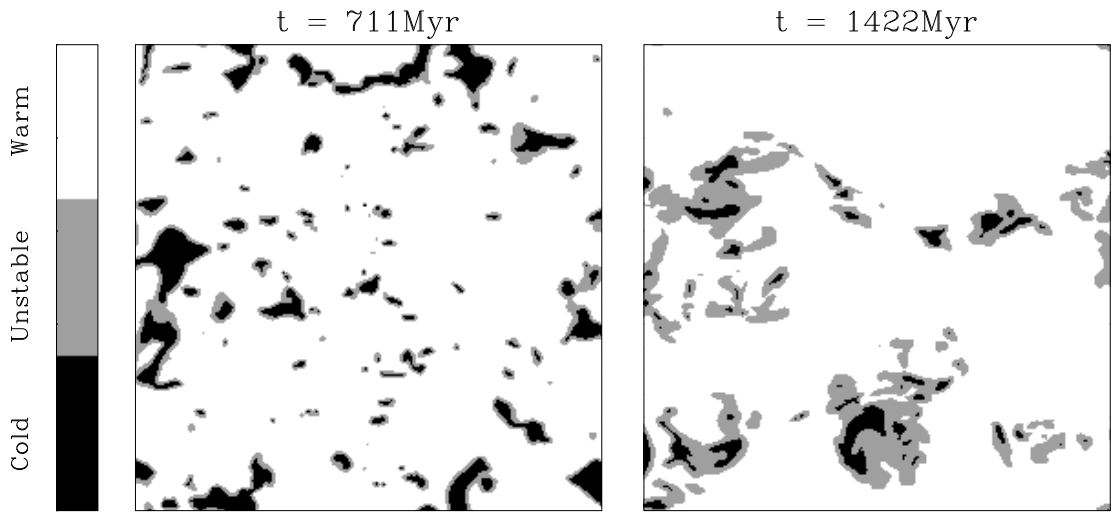}
\caption{A slice at y=constant through the data cube showing the distribution of cold, unstable, and warm gas, before and after the MRI has begun to dominate the dynamics.
\label{phaseplot}}
\end{figure}

The spatial resolution of the synthetic observations was doubled to
determine if the structure of the velocity profiles would be affected.
Generally, the line profiles remain single-peaked, whether or not
well-separated spatial components are present.  In Figure
\ref{line_x_warm} we show velocity profiles for the warm gas only, and
do not include thermal broadening.  The line-of-sight position profile
(also in Figure \ref{line_x_warm}) shows that the warm gas is
spatially much more uniformly distributed than the cold gas, which
dominates the profiles in Figures \ref{line_x} - \ref{line_z}.
Interestingly, however, the intermediate-temperature gas is always
associated with cold condensations.  This is clearly seen in Figure
\ref{phaseplot}, which shows slices through the volume both before and
after the onset of strong turbulence.

The line profiles for runs with different mean density are very
similar to the standard run.  They typically show a single component
with occasional evidence for a weaker second component.  As the mean
density is decreased to $\rm{n=0.67}$ and $0.25 \ \rm{cm^{-3}}$, the
line widths increase to 2.0, 2.2, and 1.9 $\rm{km \ s^{-1}}$, and 4.8,
4.7, and 4.5 $\rm{km \ s^{-1}}$, respectively.  Without thermal
broadening the line widths are reduced to 1.1, 1.5, and 0.74 $\kms$,
and 2.2, 2.2, and 1.1 $\kms$, respectively.  A similar trend of
decreasing line width with increasing mean density is also observed.

\section{Summary and Discussion} 
\label{discuss}

In this paper, we present results from a set of numerical MHD
simulations that focus on the interrelationship between turbulence and
thermal structure. The models we have performed are three dimensional,
and include sheared galactic rotation and magnetic fields.  Turbulence
therefore is generated by the magnetorotational instability.  We also
include a radiative cooling function that, in pressure equilibrium,
would yield a two-phase medium. The two fundamental issues we have
addressed are (1) how cloudy structure alters the saturated-state
properties and scalings of MRI-driven turbulence, compared to
single-phase MRI models, and (2) how turbulence that is {\it not}
driven by direct (stellar) thermal energy inputs affects the thermal
balance and phase structure in the warm/cold atomic medium.

\subsection{Summary of Model Results} 

Our primary findings are as follows:

1. {\it Evolution and physical structure:} A two-phase cloudy medium
with many small clouds develops in the first 20 Myr of our
simulations.  Over time, due initially to galactic shear, and later
($t>5$ orbits) to MRI-driven turbulence, these clouds undergo a
continual series of mergers and disruptions, leading to a late-time
state in which the mass function of condensations peaks at a few
hundred $M_\odot$.  The dense condensations are triaxial, and
typically have max:min axis ratios of 2:1. They consist of cold gas
lumps surrounded by envelopes of thermally-unstable gas; filling all
of the remaining volume is thermally-stable warm gas.

2. {\it Density and temperature distributions:} For the range of
parameters we have explored, in late stages of evolution most of the
gas mass is in the cold phase, while most of the volume is occupied by
the warm phase.  While the proportion of thermally-unstable
intermediate-temperature gas in a given model increases after the
advent of MRI, at all stages the density and temperature PDFs show
distinct warm and cold phases, with a varying amount of material in
the ``non-equilibrium'' valley between these peaks.  The peaks, near
$T=100\ {\rm K}$ and $T=8000\ {\rm K}$, also broaden as the turbulence
develops. The relative proportions in each phase depend on the mean
density, varying from 95\% cold gas when the mean density $\bar n=4.0\ 
{\rm cm}^{-3}$, to 50\% cold gas when $\bar n= 0.25\ {\rm cm}^{-3}$.
The fractions of thermally-unstable gas and warm gas are always
comparable to each other.  Increasing levels of turbulence yield
increasing proportions of thermally-unstable gas.  Relative to the
proportions predicted for a two-phase, quiescent medium in thermal and
pressure equilibrium, increasing turbulence also tends to {\it
  increase} the fraction of cold gas, while decreasing the fraction of
warm stable gas.

3. {\it Pressure:} We initialize our models at $P/k=2000\ \rm{K \ 
  cm^{-3}}$, but secular cooling in the stages before MRI develops
leaves the gas in approximate pressure equilibrium ($\Delta P/\bar P<
0.1$) at a lower mean pressure of $P/k=1300\ \rm{K \ cm^{-3}}$, near
the minimum for which two stable gas phases can be present.  After MRI
develops, pressures cover a much wider range of values ($\Delta P/\bar
P\sim 0.5$), with a maximum at $P/k\sim 4000\ \rm{K \ cm^{-3}}$, but
relatively unchanged mean value ($\bar P/k=1200\ \rm{K \ cm^{-3}}$).

4. {\it Turbulent velocities:} After the MRI saturates at $\sim 5$
orbits, the turbulent velocity dispersion reaches a quasi-steady
plateau -- albeit with fluctuations of $\sim 30\%$ in amplitude.  For
our fiducial model with $\bar n=1 \ {\rm cm}^{-3}$, the mean late time
(3D) velocity dispersion is $\delta v \equiv \langle v_x^2 +(\delta
v_y)^2 +v_z^2\rangle^{1/2} \approx 2.7 \ {\rm km \ s^{-1}}$ for all
three components.  This velocity corresponds to mean Mach numbers of 0.4,
1.8 and 2.9 in the warm, intermediate, and cold phases.  We examined
the effect of mean density on the velocity dispersion, and found that
$\delta v \propto \bar n^{-0.77}$ overall, with slightly shallower
slope for the warm gas alone.  Our results show, additionally, that the
Reynolds stress, $\langle \rho v_x \delta v_y \rangle$, varies with mean
density $\propto \bar n^{-1.1}$.  We find that the in-plane components
of the velocity dispersion exceed the component perpendicular to the
disk by about a factor of two.

5. {\it Magnetic fields:} For the present set of models, we have
adopted initial conditions with a uniform vertical magnetic field of
strength $0.26 \ \mu$G.  The MRI enhances the field by an order of
magnitude, so that $\langle B^2\rangle^{1/2}$ is typically $2-3 \ \mu$G
late in the simulation.  The field strength is similar (within $\sim
20\%$) in all three phases of gas, and there is no significant trend
of field strength with mean simulation density $\bar n$.  However, we
find that the Maxwell stress, $\langle -B_x B_y/(4\pi)\rangle$, varies
with mean density $\propto \bar n^{-0.4}$.

6. {\it Synthetic line profiles:} As a demonstration of the potential
for employing simulations to interpret observational diagnostics, we
have computed maps of synthetic line profiles from sample data cubes.
We find that the line profiles are generally single-peaked (although
in some cases would require two or three components if a standard
Gaussian fitting scheme were performed).  In no case did we identify
two distinct velocity components, even though there are distinct cloud
structures present along many lines of sight.  Because turbulence has
a smooth power spectrum, this kind of overlap in velocity space is
inevitable.

Our results have several interesting implications for interpreting ISM
observations, and it is also interesting to compare with recent
numerical and theoretical work on the ISM and on MRI dynamics.  We
conclude by discussing these connections.

\subsection{The Multiphase MRI and Saturated-State Turbulence}

High levels of turbulence are observed both in the atomic gas of the
Milky Way, and in that of external spiral galaxies, and it has been
suggested that the MRI could be an important contributor to this
turbulence, especially in the outer parts of galaxies where there is
little star formation.  Our models are the first (to our knowledge) to
address this issue directly with an appropriate physical model --
namely, one that admits two stable thermal phases, such that the MRI
must develop in a cloudy medium with density contrasts of 100 between
clumps and diffuse gas.  While the turbulent velocity that develops in
our fiducial model with mean density $\bar n=1 \ {\rm cm}^{-3}$ is
relatively modest, the scaling of the turbulent amplitude with $\bar
n$ is quite steep, such that $\delta v \sim 8 \ {\rm km \ s^{-1}}$
is predicted when $\bar n=0.2\ {\rm cm}^{-3}$.

The scaling of turbulent velocity dispersion with mean density
indicates that MRI may play a significant role in the outer regions of
the galaxy.  Beyond the point in the Milky Way where the stellar
surface density drops, the gas scale height rapidly increases, and the
volume density correspondingly decreases; this sort of disk flaring is
also seen in external galaxies.  In the \cite{wol03} Milky Way model,
for example, $\bar n$ falls below $0.2 \ {\rm cm}^{-3}$ at $R=15 \ 
{\rm kpc}$.  The outer-galaxy pressure in the \cite{wol03} model is
nevertheless high enough for cold-phase gas to be present, so that if
it were {\it not} turbulent, then a thin gravitationally unstable
layer would develop.\footnote{For a cold layer without turbulence, the
  value of the Toomre Q is $< 1.5$ (the threshold for instability) for
  ${\rm R} \Sigma > 12 \ {\rm kpc} \times 1 M_{\odot} \pc^{-2}$, where
  $\Sigma$ is the gas surface density.  \citet{wol03} show that cold
  gas is expected to be present at all radii out to 18 kpc.  Since
  $\Sigma$ is estimated to exceed $3 M_\odot \pc^{-2}$ inside this
  radius, even if only 20\% of the gas is in the cold component, the
  layer would be gravitationally unstable.}  Our results suggest that
the MRI could be maintaining high-amplitude turbulence, and hence
suppressing star formation, in the far outer Milky Way and other
spiral galaxies.  We note that the increase in scale height is
necessary for this to hold; since the minimum MRI wavelength $\propto
\bar n^{-1}$, MRI-driven turbulence can only be sustained in a
sufficiently thick disk.

Even in the inner disk, our results suggest that MRI may be a
significant contributor to turbulence in the ISM.  At a mean
inner-Galaxy (${\rm R < 10 \ kpc}$) midplane density of $\bar n=0.6 \ 
{\rm cm}^{-3}$ \citep{dic90b,wol03}, our results would predict $\delta
v \approx 4 \ {\rm km \ s^{-1}}$.  Away from the midplane where the
density drops, the turbulent amplitudes would increase.  The mean
inner-disk vertical magnetic field strength may also be somewhat
larger than the fiducial value we have adopted (\cite{han99} obtained
$\langle B_z\rangle=0.37 \ \mu$G from pulsar observations), which
would tend to increase the amplitude of the turbulence.  A more
extensive parameter survey -- allowing for disk stratification,
varying scale height, and differing initial field strengths and
distributions -- is needed to quantify more fully the expected
contribution from MRI to turbulent amplitudes in the ISM.  Another
important question is whether MRI development could be quantitatively
altered by interaction with large-scale pertubations driven by
supernovae or spiral shocks.  We defer consideration of this
interesting issue to future work.

Direct comparisons between simulations and observations regarding
levels of turbulence and magnetic field strength as a function of
local parameters would be very useful. Unfortunately, observations at
this time do not permit such comparisons to be made.  HI velocity
dispersions in the Milky Way can only be measured within $\sim$ 1 kpc of
the Sun (e.g. \cite{loc91,hei03}).  In external near-face-on galaxies,
observed (vertical) velocity dispersions combine both turbulent and
thermal contributions, and these values do not vary secularly with
galactic radii (although dispersions are significant) even well
beyond the optical disk (see \S 1).  Magnetic field strengths in the Milky Way
beyond $\sim$ 10 kpc have not been measured directly (i.e. with
Faraday rotation; see e.g. \citet{han04}).  For both the Milky Way and
external galaxies, one may use synchrotron emission \citep{bec04} to
obtain the product of the magnetic and cosmic ray energy densities as
a function of galactic radius, but since the equipartition assumption
need not be satisfied everywhere, this does not yield a B-field
strength except locally, where electron cosmic-ray and gamma-ray
observations can be made.  Milky Way outer-galaxy field strengths of
2-3 $\mu$ G are consistent with synchrotron/cosmic ray models of
\citet{str00}.

The scalings we find for MRI amplitudes show interesting differences
from those obtained with single-phase gaseous media, in previous
adiabatic and isothermal simulations.  In the shearing-box models of
HGB1, HGB2, and \citet{san04}, all of the measures
of turbulence scale together -- i.e. $\langle B^2 \rangle \propto
\langle -B_x B_y \rangle \propto \langle \rho (\delta v)^2 \rangle
\propto \langle \rho v_x \delta v_y \rangle $.  For simulations with
net vertical magnetic flux, HGB1 further reported that these stresses
and energy densities scale $\propto L_z \Omega v_{A, z}/c_s^2$, where
$v_{A, z}$ and $c_s$ are the initial Alfv\'en speed and sound speed,
respectively.  \citet{san04}, on the other hand, identify a scaling of
stresses $\propto v_{A, z}^{3/2}$ in the mean Alfv\'en speed, and also
demonstrate that the pressure dependence of the saturated-state
stresses are very weak.

While we have not surveyed cases with differing magnetic field
strengths, our models at varying mean density have varying $v_{A, z}$.
At fixed mean $B_z$, the single-phase medium simulations cited above
would predict scalings $\propto \bar n ^{-1/2}$ or $\propto \bar n
^{-3/4}$ for $\langle B^2 \rangle$, $\langle \rho (\delta v)^2
\rangle$, and the stresses.  For our cloudy-medium simulations, we in
fact find that $\langle B^2 \rangle$ is nearly independent of $\bar
n$, while other scalings in our models are either in the same range as
the single-phase predictions (i.e. $\langle \rho (\delta v)^2 \rangle
\propto \bar n^{-0.54}$ and $\langle -B_x B_y\rangle \propto \bar
n^{-0.4}$ ), or slightly steeper ($\langle \rho v_x \delta v_y \rangle
\propto \bar n^{-1.1}$).  Interestingly, 3D MRI simulations in
radiation-dominated disks \citep{tur03}, which like our models contain
strong density contrasts, also show a steep dependence of the stress
$\propto v_{A,z}^2$.

The reason for the difference between single-phase and cloudy-medium
results for saturated-state $\langle B^2 \rangle$ scalings is not yet
clear, but the lack of dependence on $\bar n$ in our models suggests
that local, rather than global, properties of the gas determine the
field strength that develops.  The densities in the diffuse phase and
in the dense phase are similar for all our models; only the filling
factor of cold clouds differs appreciably.  If the saturated-state
field strength depends on the reconnection rate, and this depends on
local gas densities and field geometry, then the fact that these
properties are the same in all our two-phase models might explain the
lack of dependence of $\langle B^2 \rangle$ on $\bar n$.  Our
isothermal model, which has less-kinked magnetic field, and
typical (log) densities midway between $\log(n_{warm})\sim -1$ and
$\log(n_{cold})\sim 1$, indeed has saturated-state values of the RMS
field strength 20-50\% higher than the standard run ($3.5 \ \mu$G in the
isothermal model, compared to 2.4, 2.7, and 2.9 $\mu$G in the F, G,
and H phases for the standard run).  Further study, with particular
focus on the rate and spatial distribution of reconnection, is needed
to clarify this issue.

While the saturated-state magnetic field strength depends on a balance
between MRI-driven amplification and (numerical or true resistive)
dissipation, the saturated-state velocity dispersion depends on a
balance between MRI driving and losses in shocks, compressions, and
(numerical or true viscous) shear dissipation.  Since turbulent
velocities are similar in all the gas components, while the mass is
concentrated in the cold clouds, in Paper I we proposed that the cloud
collision time may be a good proxy for the kinetic energy dissipation
time. The kinetic energy dissipated per unit time per unit mass is
then $\dot {\cal E}_{diss}\sim (\delta v)^3 \bar \rho/(r_{cl}
\rho_{cl})$, where $r_{cl}$ is a cloud size and $\rho_{cl}\approx
\rho_{cold}$ is the gas density within clouds.  The kinetic energy
input rate per unit mass due to MRI is an order-unity constant times
$\Omega \langle -B_x B_y \rangle/(4\pi \bar \rho)$.  Balancing inputs
with dissipation, and using $\langle -B_x B_y \rangle \propto \bar
\rho^{-0.4}$ from our simulations, this predicts a scaling for the
saturated-state velocity dispersion $\delta v \propto \bar \rho
^{-0.8}$.  This prediction indeed agrees well with the velocity
dispersion scaling measured directly from our simulations, $\delta v
\propto \bar \rho ^{-0.77}$.  Of course, this scaling cannot continue
to arbitrarily low density, because the MRI becomes stabilized if the
wavelength $(\propto 1/\sqrt{\rho})$ exceeds the height of the disk.

In previous analytic work on magnetized cloud-cloud collisions
\citet{cli83} argued that effective cloud cross sections should vary
$\propto (\bar \rho \delta v)^{-2/3}$. Although we do not measure this
effect directly, it would not be expected to change our results
significantly.  Taking this into account for $r_{cl}$ in the above
analysis, the predicted power-law scaling exponent for $\delta v$ with
$\bar \rho$ changes by only 0.02.

\subsection{Structure and Thermodynamics of the Atomic ISM }

As noted above, our results show only minor differences between
typical magnetic field strengths in gas of different phases, with the
cold medium having slightly higher $B$.  If flux were frozen to the
gas, and cold clouds contracted isotropically out of the warm ISM due
to TI, then the field strengths would differ by a factor 20 between
diffuse and dense phases.  Evidently, however, this is not the case:
observationally, \citet{hei04} reports the $B$-fields in cold
atomic clouds are no stronger than in the diffuse ISM overall.  Our
models are consistent with this result, in part because clouds in fact
do not condense isotropically (but instead preferentially gather
material along field lines), and in part because field lines can
diffuse relative to matter (in our models, this is purely numerical,
but turbulent diffusion is likely to play the same physical role).

Synthetic line profiles of the sort we have computed are potentially
of great value in interpreting observations of the ISM, and in
particular, the emission and absorption profiles of the 21 cm hydrogen
line.  Ensemble properties of the atomic ISM, including the turbulent
velocity dispersion for the cold gas separately, can be unambiguously
obtained from observed position-velocity data cubes.  Our results
illustrate the difficulty, however, in discerning detailed (local)
structural properties of the ISM directly from line profiles, due to
overlapping in velocity space (cf. Ostriker, Stone, \& Gammie 2001).
Fortunately, recent work has demonstrated that analysis techniques
calibrated using simulations can be used to extract fundamental
statistical properties such as power spectra from CO observations of
turbulent molecular clouds (e.g. \cite{bru03, hey04}), and it will be
interesting to test whether the same holds true for HI gas.

What do we conclude regarding the dynamics/thermodynamics connection
in the turbulent atomic ISM?  As alluded to in \S 1, recent
observations \citep{hei03} have suggested that at least 30\% of the HI
gas at high latitudes (where blending due to galactic rotation is
minimized) is in the thermally-unstable regime ($T=500-5000$ K).
Another 30\% of the HI mass at high latitudes is estimated to be warm,
thermally-stable gas.\footnote{At low latitudes, however, there is a
  very strong peak in the total gas column at $T\approx 8000-9000$K in
  the \citet{hei03} data, suggesting that midplane gas may be much
  more likely to be in thermal equilibrium.}  Interestingly, we find
that in our simulations, the warm and thermally-unstable gas mass
fractions are also always very similar.  In our models, the
warm+unstable mass fraction only approaches 50\% for the lowest-$\bar
n$ case; at $\bar n=0.67 {\rm cm}^{-3}$, the warm+unstable gas
comprises 30\% of the total mass.  However, except at the lowest $\bar
n$, the turbulent amplitudes in our models are also somewhat lower
than in the real ISM, so in part the lower warm+unstable fractions we
find may owe to lower rates of turbulent dissipation.  In addition, as
discussed in \S 3.6, because our numerical method is non-conservative,
a significant fraction of the input turbulent energy from the MRI is
lost without being thermalized.  Our method therefore underestimates
the heating rate.  This may be especially true in high density regions
where the magnetic field curvature is large, so that the current
density ${\bf J}=\nabla\times{\bf B}$ is large, and the resistive
heating $\propto |{\bf J}|^2$ should also be large.  We intend to
explore this issue in future work incorporating explicit resistivity,
also comparing with results using a conservative algorithm to update
the energy equation.

Overall, our results are consistent with the recent work of
\citet{aud04}, who performed high resolution 2D simulations of a
converging flow in which turbulence develops, using a cooling function
very similar to ours.  The temperature and density PDFs from their
models are qualitatively similar to ours, in particular showing
evidence for the existence of a two-phase medium even in the most
turbulent case.  Similar to our results, they found that the
proportion of thermally-unstable gas increases with the amplitude of
turbulence. For their least turbulent model, about 10\% of the gas was
thermally unstable, while this fraction increased to 30\% in models
with larger turbulent amplitudes.

Other recent work, based on simulations that include modeled effects
of star formation via prescriptions for OB star heating (e.g.
\citet{vaz00,gaz01,vaz03}) and supernovae (e.g.
\citet{ros95,kor99,dea00,wad00,wad01a,wad01b,bal04,mac04}), have found
significant fractions of gas in the unstable phase, and many have advocated
for a shift away from the classical two- or three-phase medium concept
towards more of a ``phase continuum'' description.  However, the
former set of models have energy inputs from ``star formation'' in
cold atomic gas (rather than in self-gravitating molecular clouds),
which likely leads to overestimating the amount of thermally-unstable gas.
The latter models have not included low-temperature cooling, so no
cold phase can form.  Thus, it is not yet clear whether realistic
models incorporating turbulence driven by star formation would indeed
produce an extended, featureless continuum of temperatures, or whether
they would yield bimodal distributions similar to those we have found
with solely MRI-driven turbulence.

We believe that, given the observed turbulent amplitudes ($\sigma_v
\sim 10 \ {\rm km \ s^{-1}}$) and characteristic spatial scales (the
disk thickness $H\sim 100$ pc) for the atomic ISM, quasi-two-phase
distributions are inevitable. The heating time $\sim H/\sigma_v =
10^7$ yr is a factor of a few longer than the cooling time in gas at
densities $\simlt 1 {\rm cm}^{-3}$ (see eq. [5] of Paper I or eq. [4]
of \citet{wol03}), so that moderate but not extreme departures from
thermal equilibrium can be expected in diffuse gas.  In dense gas,
cooling times are very short, so thermal equilibrium must hold, but
typical order-unity variations in pressure from sonic-level
turbulence can lead to moderate local density and temperature
variations.  Future work will be able to determine whether, with
realistic turbulent amplitudes and fully-captured energy dissipation,
thermal distributions from models can indeed match those from HI
observations, or whether additional heat sources are required.

\acknowledgments

We are grateful to Charles Gammie, Woong-Tae Kim, Jim Stone, and Mark
Wolfire for valuable discussions, and Yen-Ting Lin and Jonathan
McKinney for helpful technical advice.  The manuscript has also
benefited from a thoughtful and detailed report prepared by an
anonymous referee.  This work was supported in part by grants NAG
59167 (NASA) and AST 0205972 (NSF).  Some of the computations were
performed on the Tungsten cluster at NCSA, and others were performed
on the CTC cluster in the UMD Department of Astronomy.

\clearpage

\clearpage 

\clearpage 

\clearpage

\clearpage 

\clearpage 

\clearpage 

\clearpage 

\clearpage 

\clearpage 

\clearpage 

\clearpage 

\clearpage 

\clearpage 

\clearpage 

\clearpage 

\clearpage 

\clearpage

\clearpage 

\clearpage 

\clearpage 

\clearpage 

\clearpage

\end{document}